# Continuous motion of an electrically actuated water droplet over a PDMS-coated surface

Supriya Upadhyay[1], K. Muralidhar[1*]

[1]Department of Mechanical Engineering, Indian Institute of Technology Kanpur, Kanpur 208016, UP, India

*Corresponding author. E-mail: kmurli@iitk.ac.in

16-digit ORCID, Supriya Upadhyay: 0000-0003-2728-7972, K. Muralidhar: 0000-0002-8514-4323

## Abstract

Electrically actuated continuous motion of a water droplet over PDMS-coated single active electrode is analyzed from detailed modeling and experiments. In an experiment, continuous motion of the droplet is achieved when it is located over an active electrode with a horizontal ground wire placed just above in an open-EWOD configuration. Using a CCD camera, the instantaneous centroid position of the droplet is determined and its velocity is inferred by numerical differentiation. The edge-detected image is also used to determine the advancing and receding contact angles of the moving drop relative to the substrate. Motion of 2, 6, and 10 µl water droplets for voltages in the range of 170-270 $V_{DC}$ is examined to investigate the effect of drop volume and voltage on drop deformation and velocity. Simulations have been carried out using COMSOL© Multiphysics with full coupling between the electric field and hydrodynamics. The motion of the droplet is initiated by Young-Lippmann spreading at the three-phase contact line, followed by a nonuniform electric force field distributed between the active electrode and the ground wire localized at the droplet-air interface. The solver evaluates the Maxwell's stress tensor and introduces it as a volumetric electrostatic force in the Navier-Stokes equations. The fully coupled numerical solution shows a good match with experimentally determined drop movement over a silicone oil-coated PDMS layer for which contact line friction is absent. A contact angle model with friction leads to close agreement between simulations and drop motion over a bare PDMS layer. Over both surfaces, continuous motion of the water droplet is seen to be achieved in three stages, namely, initial spreading, acceleration, and attainment of constant speed. Numerical modeling that includes electric field-fluid flow coupling is shown to yield data in conformity with experiments.

# 1 Introduction

Electrowetting and actuation phenomena relate to the movement of small volumes of liquid over a surface by using an externally applied voltage. The process involves electrostatic forces on interfaces of conducting liquids with the ambient and over a three-phase contact line. The shape and motion of a liquid-gas meniscus are controlled by creating a voltage difference between the electrode and the droplet. Electrowetting-based devices could not be illustrated earlier in water since electrolysis occurs at a higher voltage, thus, limiting its operation. The notion of using a thin layer of insulating material between the conductive liquid and the metal electrode to separate them to eliminate electrolysis is known as Electrowetting on Dielectric (EWOD) (Berge 1993; Quilliet and Berge 2001) and has applications in microfluidic devices such as biomedical diagnostic systems.

Washizu (1998) demonstrated motion, sorting, and mixing of the droplet through electrostatic field actuation. These operations were performed using a device that is constructed as an array of electrodes. As a result, the electric field lines between the active electrodes and Maxwell's stresses generated by the induced charges led to droplet motion towards the active electrode. Thus, aqueous droplets were transported continuously by sequential voltage application to the electrodes in the array. Pollack et al. (2000) achieved significant improvement in conductive liquid droplet polarization and speed by applying voltage in an EWOD device. The droplet was sandwiched between a top and bottom plate. The ground connection was provided to the top electrode and a linear array of electrically actuated electrodes acted as a base plate. Ren et al. (2002) formulated a model to explain the dynamics of transportation of a liquid droplet using EWOD. The velocity of the droplet was derived as a function of the applied voltage. The viscosity of the droplet and that of the surrounding medium, contact line friction, geometry, and interfacial tension were the operating parameters. It was revealed that oil as the surrounding medium decreases the contact angle hysteresis, ultimately decreasing the threshold voltage. Lee et al. (2002) analyzed electrowetting (EW) and EWOD principles applied to microfluidic devices to control the wettability of liquids on a solid surface under the effect of an electric potential. Moon et al. (2002) examined the effect of thickness and relative permittivity of a dielectric material on the minimum actuation voltage in the EWOD system. It was experimentally confirmed that electro-wettability, that is lowering of the contact angle increased at lower dielectric thickness. Pollack et al. (2002) demonstrated the motion of 100 mM KCl droplets at 60 $V_{DC}$ at a speed of 100 mm/s by using low-viscosity (0.001 Pa-s) silicone oil as the surrounding medium. Using an immiscible liquid as a surrounding medium averts evaporation of the drop, decreases the threshold voltage by reducing the contact angle hysteresis, and lowers wall friction. Gunji and Washizu (2005) described electric field-driven self-propelling of a water droplet on hydrophobic coated co-planar-strip electrodes. At a DC voltage or single-phase low-frequency AC voltage, the droplet was seen to move along the electrodes at speed as high as 100 mm/s. The water droplet left a film along the path travelled, producing an imbalance in the Maxwell's stresses due to the shielding of the electric field on its trailing side, contributing to further acceleration. Berthier et al. (2007) performed an experiment on the motion of a droplet placed over a non-active electrode to the active electrode when a ground wire connection was passed over the droplet. On the basis of the electrocapillary force, the authors calculated analytically the maximum and minimum actuation potential for the EWOD microsystem. The minimum threshold was found to depend on the contact angle hysteresis. Baird et al. (2007) investigated the force distribution within the drop and force variation at the interface of the droplet in an EWOD device. The charge densities and potential distribution at the droplet interface were numerically determined. Arzpeyma et al. (2008) numerically studied EWOD-driven droplet motion by coupling the hydrodynamic and electrostatic forces. The change in contact angle arising from the applied voltage was calculated by the Young-Lippmann equation, while a Laplace equation was solved for determining the electric potential. Moreover, the solution of Navier-Stokes equations provided the velocity distribution and shape of the droplet. The outcome of simulation was validated against experiments. Raghavan et al. (2009) presented an experimental and theoretical analysis of electro-hydrodynamic flow produced by electrical pressure gradients. Brassard et al. (2008) experimentally performed electrical manipulation of water droplets in air and silicone oil as a surrounding medium. Between zero and 120 V, an intricate hysteresis pattern was seen in air. In contrast, almost no hysteresis was observed in silicone oil. Young and Mohseni (2008) investigated the force distribution characteristics of dielectrophoresis (DEP) and electrowetting on dielectric (EWOD), two basic transportation approaches in microfluidics. Bavie're et al. (2008) used a setup design in which a ground wire passed through the droplet. The authors used a silicone oil film as the hydrophobic layer of the chip. No significant effect of silicone oil was observed in the contact angle. Fan et al. (2010) demonstrated polarization of a neutral particle dispersed in water droplets by an EWOD parallel plate system by applying an electric field within an acceptable frequency range. A particle chain was formed due to the attraction of individual particles with induced dipole moments. Thus, electrowetting of the droplet and polarization of particles dispersed in it were demonstrated. Roghair et al. (2015) numerically determined the deformation of the interface and motion of the three-phase contact line in the presence of an electric field. The authors coupled the electrostatic force to the Navier-Stokes equations and solved the electro-hydrodynamic problem using OpenFOAM. A Cox–Voinov boundary condition was prescribed as a dynamic contact angle model for the moving three-phase contact line. Cavalli et al. (2015) introduced a three-dimensional computational approach to understanding the interaction of sessile drops with tunable wetting defects. The model was implemented by using OpenFOAM using a diffuse interface approach and a Cox-Voinov dynamic contact angle model. Li et al. (2022) determined forces experienced by a moving droplet when it slid down over a tilted surface by imaging the drop trajectory and applying relevant equations of motion. The motion of the droplet over a low permittivity substrate was found to be influenced considerably by electrostatic forces. Charging of drops was detected on all hydrophobic $SiO_2$ surfaces. In contrast, charging was at least ten times lower for the PS-on-gold, Teflon-on-gold and PDMS-on-Si samples.

Gas bubbles will be observed due to electrolysis if the dielectric fails to provide insulation between water and the electrode. Mugele and Heikenfeld (2019) discussed how dielectric leakage cause a DC current to flow and cause a portion of the applied potential to drop partially or fully across electrically resistive materials in the pathway. PDMS plays a useful role in preventing electrolysis of water, being hydrophobic but not super-hydrophobic as required in microfluidic applications. In addition, it has large hysteresis that may result in pinning. PDMS is however, well-characterized in the literature and has often been used as a reference substrate (Caputo et al., 2014; Sohail et al. 2014). Thus, PDMS becomes an essential aspect of an electrically actuated droplet. In a device, PDMS may be used with additional coatings or texturing to improve its wettability.

The conventional method to electrically transport a droplet from a non-actuating to an actuating electrode is by placing the droplet over the shared boundary of the two electrodes. The electrostatic force distributed over the three-phase contact line (TCL) falling on the actuated electrode creates an unbalanced force component that tends to pull the droplet over it. In contrast, an open electrowetting-on-dielectric (EWOD) configuration is adopted in the present study with an active base electrode and a ground wire placed horizontally above but touching the drop. Continuous motion of the droplet is achieved by using a single active electrode at a voltage that is spatially and temporally constant. Fully coupled electro-hydrodynamic simulations have been carried out in 2D as well as an axisymmetric coordinate system using COMSOL© Multiphysics. A dynamic contact angle model that includes electrowetting, friction and hysteresis is prescribed at the three-phase contact line. Such a detailed approach for continuous motion of a droplet has not been reported in the literature. Using experiments as well as simulation, shapes of a drop of DI water and the centroid velocity during spreading as well as continuous motion are compared in the present study. The response of the droplet moving over a bare PDMS surface is compared with another moving over a silicone oil-coated PDMS surface. Quantities of interest are the changes in the dynamic contact angle, contact line velocity, and the emerging relationship between the contact angle and the capillary number. The trends in drop deformation and movement seen are explained in terms of the distribution of pressure, electric field, and velocity vectors.

## 2 Experimental apparatus

Electrically driven drop motion characteristics have been determined in the present work via an extensive set of experiments. An experimental set-up for continuous imaging of motion of the droplet under the effect of electric field lines using an open-EWOD device was designed and fabricated as a part of the study. Figure 1 shows a schematic drawing of the experimental set-up in which droplet actuation experiments were carried out and is further elaborated below. A detailed description of the apparatus is available in the literature (Jain et al., 2022).

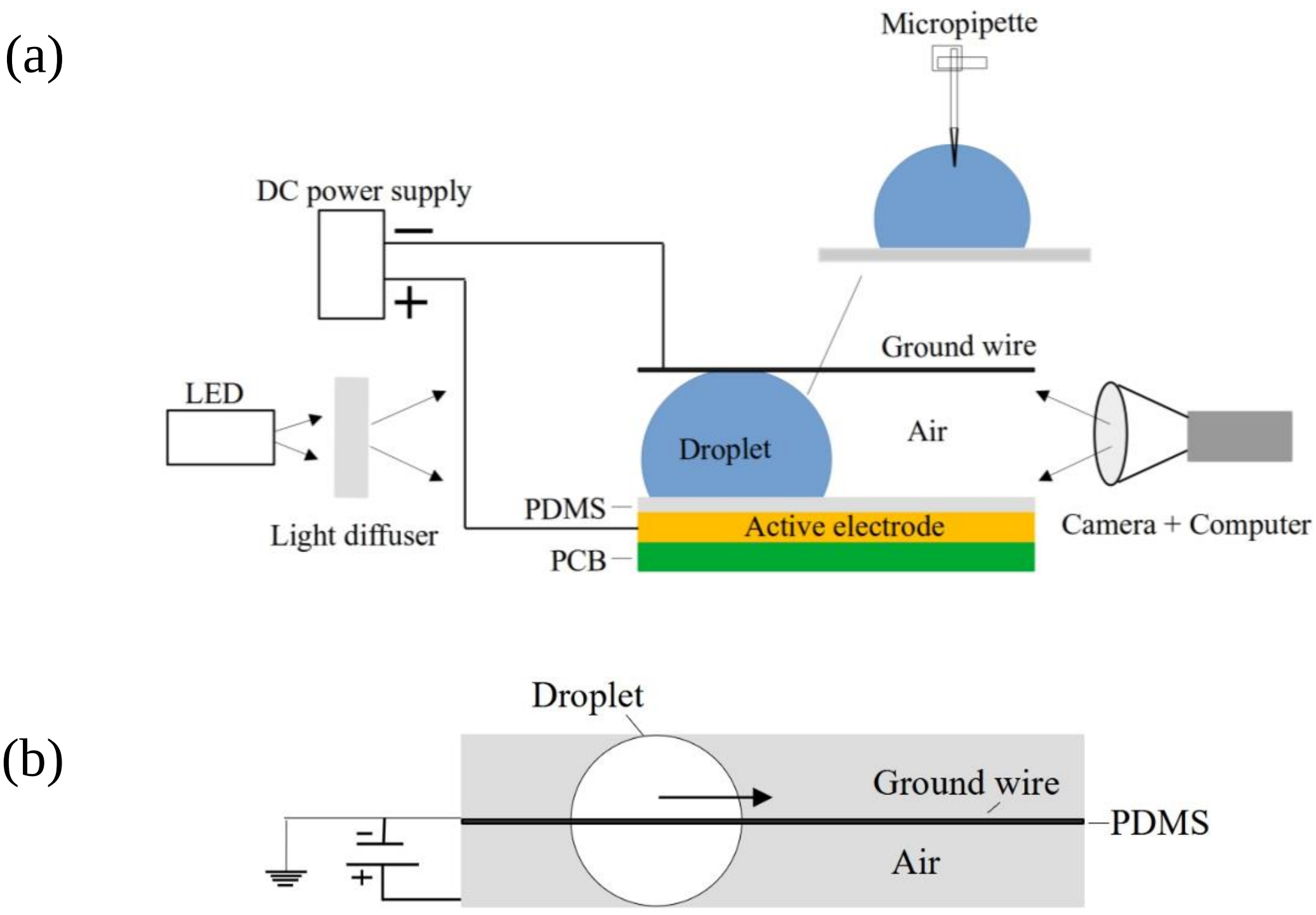

**Figure 1** Schematic drawing of the experimental set-up of an open-EWOD configuration that enables continuous motion of a droplet in the presence of an electric field (a) side-view and (b) top-view.

A video camera (Genie Nano C1280) was used to record the dynamic behavior of the droplet movement. Images were captured at 120 fps for a total of 3000 frames at 1028×1024 resolution per frame with 10-bit (RGB) digitization of light intensity. In addition, an HR F2.8 / 50 mm (far/near) (AX2 TV EXTENDER) lens was used with the camera for a magnified view of the drop. In the present experiments, the scale factor was 8.33 μm per pixel for the given position of the camera relative to the midplane of the drop. The apparatus was built on an acrylic stage with a white light LED source and a light diffuser for proper illumination of the droplet interface. In addition, it included a portable substrate holder, micropipette, and a voltage control unit. A white light LED source with a light diffuser delivered proper illumination of the droplet interface with the ambient. Aluminium wire, 0.1 mm diameter was used as a ground wire electrode. The PCB carrying the electrode was placed over an acrylic stage that has two adjustable banana jacks on each side. The ground wire was soldered to the banana jacks and passed along the centreline of the lower electrode pad while maintaining a uniform gap with the surface. While the alignment of the wire was optically performed (± 0.2°), it was held taut during the duration of the experiment. A direct current power supply unit was used to activate the electrode. In the presence of the ground wire, a potential difference is created between the droplet and the electrode. Since current flow in the apparatus is negligible, Joule heating is negligible in the study (Section 5).

## 2. 1 Surface preparation

The EWOD device is developed over printed circuit board (PCB) and a dielectric coating of PDMS over the copper electrode also serves as a hydrophobic layer. Experiments were carried out with and without silicone oil coating over the PDMS-coated PCB substrate. These correspond respectively to a sticky hydrophobic surface (bare PDMS) and a lubricated hydrophobic surface (silicone oil coating over PDMS). The PCB was designed using Autodesk EAGLE software and procured from a commercial vendor (PCB Power Market, India). The present work has a single long electrode pad with dimensions of 80 mm × 4 mm.

PDMS coating over the PCB is obtained using a spin coater. PDMS pre-polymer in a 10:1 ratio of base to curing agent is mixed well for 10 min and placed in a vacuum desiccator for 10 min to remove air bubbles. PDMS is spin-coated (2000 rpm, 65 s) on the patterned PCB substrate. The device is then cured at 100°C for 1 hour, yielding a PDMS layer thickness of 35 μm. Using a goniometer, bare PDMS was found to have an average equilibrium contact angle of 104°. In addition, a silicone oil layer ($\mu$ = 5 mPa-s) was coated over the cured PDMS layer by spin-coating at 4000 rpm. The contact angle remained unchanged for this surface as well. The top wire that serves as the electrical ground has a diameter of 100 μm. A potential difference is created between the droplet and the electrode surface by applying the ground voltage to the top wire and a positive value to the lower electrode. The water droplet is dispensed over the surface using a micropipette. A fresh surface was adopted for a new experiment so that the previously formed liquid film did not play a role. Experiments on electrical actuation is initiated after the initial drop velocities are fully damped over a time frame of around 10 s.

# 3 Experimental data analysis

The recorded videos were converted into an individual sequence of images for data analysis. The edge of the droplet was determined by using the image processing tools of MATLAB. A canny edge detection approach was used to detect the edge of the droplet. Noise content was low throughout the experiments and the correct edge of the droplet was detected without ambiguity. A sample video has images of a 6 µl droplet under static and moving conditions (see video in the supplementary material). The images are recorded in JPG format, and the image data is available in the form of light intensity on a 0-255 scale at each of the pixels. The important steps in data analysis are edge detection and contrast improvement, apart from calculating the dynamic contact angle and the coordinates of the droplet centroid.

## 3.1 Measurement of advancing and receding contact angles

The contact angle of the droplet is found by evaluating the tangent of the angle in the form of the slope of the local line passing over the droplet surface. The tangent calculation proceeds towards the solid wall but terminates a few pixels away from it. The collection of tangent values is then extrapolated to the wall to yield $\tan\ \theta_{eq}$ for a static drop and $\tan\ \theta_d$ for a moving drop. These respectively determine values of the equilibrium and dynamic contact angles under electrowetting conditions. The entire calculation for a single frame and a video can be carried out within the MATLAB© software.

## 3.2 Locating centroid of the droplet

The coordinate values of each pixel in the two directions of the image plane can be found using an appropriate scale factor. The centroid is then determined using a weighted mean approach as per the following expressions:

$$x_c = \frac{1}{W}\sum_{i=1}^{N} x_i \cdot w_i \ ; \quad y_c = \frac{1}{W}\sum_{i=1}^{N} y_i \cdot w_i \tag{1}$$

Here, $W = \sum_{i=1}^{N} w_i$ and $i = 1...N$ is a pixel index with planar coordinates $(x_i, y_i)$; in addition, $w_i$ = 1 is the weight factor for pixels within the drop boundary and $w_i$ = 0 for pixels outside the droplet. In most imaging experiments, $N = 1028 \times 1028 \sim 10^6$.

## 3.3 Determination of the centroid velocity of the moving droplet

Centroid calculation enables the determination of the area-representative velocity of the droplet. Here, centroids calculated from adjacent frames of the video sequence are assembled, and their positions are differentiated numerically to obtain the velocity components; details of the frame-wise calculation can be seen in Somwanshi et al. (2022) and Upadhyay (2021). Centroids from three adjacent frames have been used to determine one pair of velocity components in the present study. The video sequence referred here was recorded at 120 frames per second, yielding a time interval between adjacent frames of 1/120$^{th}$ of a second.  The horizontal component velocities measured in the present study are in the range of 0.24-1.7 mm/s, while the vertical component of velocity is smaller by two orders of magnitude.

## 4 Mathematical modeling

A model of droplet spreading and continuous motion over a single active electrode using COMSOL Multiphysics simulation software is developed. Modeling of droplet spreading is performed on an axisymmetric geometry while continuous motion of the droplet is simulated using a two-dimensional Cartesian coordinate system. The continuous motion of the droplet involves initial spreading, drop deformation, acceleration, and motion at a constant speed. Hence, droplet motion needs to be modeled by considering appropriate length and timescales. A mathematical model of spreading and motion of a droplet over an active electrode in the presence of an electric field is described in the present section.

### 4.1 Navier–Stokes equations and treatment of source terms

The Navier–Stokes equations and the incompressibility constraint that model fluid motion are given as:

$$\nabla \cdot u = 0 \tag{2}$$

$$\rho\left[\frac{\partial u}{\partial t} + \rho\left(u \cdot \nabla\right)u\right] = -\nabla p + \nabla \cdot \mu\left(\nabla u + \left(\nabla u\right)^{T}\right) + F_G + F_{ST} + F \tag{3}$$

Here $u$ is the velocity vector, $F_G$ is the force due to gravity, $F_{ST}$ is the interfacial surface tension, and $F$ is a user-defined body force, namely, the volumetric electrostatic force in the present context.

### 4.2 Electro-hydrodynamics modeling

The present study determines changes in drop shape and motion driven by an electric field starting from equilibrium, followed by acceleration under electrical actuation and finally, uniform motion. The horizontal component of the electrostatic force produces motion of the droplet in the forward direction over the substrate that affects the shape as well. The fluid flow module is thus coupled with the electrostatic module within the simulator.

The electrostatic forces originate from the Maxwell's stress tensor that depends on the electric field distribution and is given as:

$$\overleftrightarrow{T} = \begin{pmatrix} \varepsilon_o\varepsilon_r E_x^2 - \frac{1}{2}\varepsilon_o\varepsilon_r\left(E_x^2 + E_y^2\right) & \varepsilon_o\varepsilon_r E_x E_y \\ \varepsilon_o\varepsilon_r E_x E_y & \varepsilon_o\varepsilon_r E_y^2 - \frac{1}{2}\varepsilon_o\varepsilon_r\left(E_x^2 + E_y^2\right) \end{pmatrix} \tag{4}$$

In the present model, the potential difference is created between the droplet and the electrode with the help of the ground wire. The computational fluid dynamics (CFD) solver is coupled with the electrostatic module (ES) by introducing the electrostatic body force into the incompressible Navier-Stokes equations. The solver evaluates the Maxwell's stress tensor first to determine the volumetric electrostatic force. Since two different fluid phases (water and air) are involved, fluid motion and the electrostatic field are further coupled via relative permittivity using the phase-field variable $\phi$ as:

$$\varepsilon_r = \varepsilon_w + \left(\varepsilon_a - \varepsilon_w\right)\phi \tag{5}$$

Here, $\varepsilon_w$ and $\varepsilon_a$ are the relative permittivity of water and air, respectively. The body force components originate from the Maxwell's stress tensor that depends on the electric field distribution and are given as:

$$F_x = \left(\frac{\partial T_{xx}}{\partial x} + \frac{\partial T_{xy}}{\partial y}\right) \qquad F_y = \left(\frac{\partial T_{yx}}{\partial x} + \frac{\partial T_{yy}}{\partial y}\right) \tag{6}$$

## 4.3 Phase-field modelling

Singularity in velocity at the three-phase contact line induces a sharp moving two-phase interface separated by zero flux in a contact line model of a partially wetting droplet on a solid surface. In a phase-field model, no-slip boundary conditions are used at the contact line without any difficulty, as it considers a diffuse interface.

The phase-field model is based on the Cahn-Hilliard equation, which has been successfully utilized in various multiphase flow models, including the wetting phenomenon of droplets over surfaces (Jacqmin 1999, 2000; Yue et al. 2004). The main advantage of using the phase-field method is that it naturally allows the motion of the contact line of the droplet along the wetted wall without any ad hoc treatment for stress singularity. Therefore, the phase-field model is favored for a simulation involving contact-line motion on a wetted surface. Here, a finite thickness of the diffuse interface separates the two phases of the system. Random molecular motion and inter-molecular attraction determine the thickness of the interface. The phase-field function governs the evolution of the interfacial layer through the Cahn-Hilliard equation.

Air and water constitute the two phases of the material region. Their thermophysical properties are evaluated at 25ºC and are available within the simulator. The Cahn-Hilliard equation that gives rise to the phase-field method is presented as

$$\frac{\partial \phi}{\partial t} + u \cdot \nabla \phi = \nabla \cdot \frac{\gamma \lambda}{\varepsilon_{pf}} \nabla \psi \tag{7}$$

Here $\phi$ is a phase variable that varies over [-1,1]. Specifically, $\phi = 1$ for water and $\phi = -1$ for air.

The fluid properties vary with the phase variable, with a jump at the gas-liquid interface. Hence, the expressions for density and viscosity at any point in the physical domain are given as:

$$\rho = \rho_a v_{f,1} + \rho_w v_{f,2} \tag{8}$$

$$\mu = \mu_a v_{f,1} + \mu_w v_{f,2} \tag{9}$$

Here subscript *a* and *w* donate air and water phases, respectively. In addition, the volume fraction distribution of the individual fluids anywhere in the fluid region is calculated as:

$$v_{f,1} = \frac{1-\phi}{2} \text{ and } v_{f,2} = \frac{1+\phi}{2} \tag{10}$$

leading to

$$v_{f,1} + v_{f,2} = 1 \tag{11}$$

The phase-field model incorporates a mobility tuning parameter, used to provide numerical stability. The mobility tuning parameter plays a crucial role in controlling the size and shape of the droplet and hence, the motion of the interface boundary. In addition, the mobility tuning parameter determines the timescale of diffusion and relaxation time of the interface in the Cahn-Hilliard equation (Yue et al. 2004). This criterion is applied for estimating the interface thickness and correctly capture the physical process of droplet deformation and motion. In the present work, simulations have been performed at various values of the mobility tuning parameter. In drop spreading simulation, the default value of MTP is employed. In the range of 70-120 m-s/kg, it is found to yield a good match with experiments for continuous motion of the droplet.

## 4.4 Dynamic contact angle modeling

The present simulation of the contact angle for a wetted wall boundary condition at the three-phase contact line uses the Young-Lippmann (Y-L) equation for the static contact angle. The dynamic contact angle model is adopted from the literature (Oh et al. 2010; Annapragada et al. 2011; Lu et al. 2017) to impose the effect of contact line friction (CLF) and pinning on droplet movement. At every time step, the velocity component of the contact line is used, which subsequently defines the dynamic contact angle. Furthermore, the numerical simulation was validated for drop spreading using a dynamic contact angle model based on the molecular kinetic theory (MKT) (Section 4.7). The Y-L equation is obtained by a force balance at the three-phase contact line (TPCL) that combines components of surface tension and electrostatic forces. The drop is assumed to relax from its initial equilibrium contact angle to the Y-L value over a time period of a few microseconds while simulations continue for over several milliseconds. The Y-L equation, molecular kinetic theory (MKT) model with CLF, and a DCA model with CLF and pinning are given as follows:

**Young-Lippmann equation (**Lu et al. 2017)

$$\cos\theta_{YL} = \cos\theta_{eq} + \frac{\varepsilon_o \varepsilon_d}{2d\sigma_{lv}} V^2 \tag{12}$$

**Molecular kinetic theory model with contact line friction (**Annapragada et al. 2011)

$$\cos\theta_{e,dyn} = \cos\theta_{eq} + \frac{1}{\sigma}\left(\frac{k\varepsilon_o \mathrm{V}^2}{d} - \xi v_{CL}(t)\right) \tag{13}$$

**Dynamic contact angle with contact line friction and pinning** (Oh et al. 2010; Annapragada et al. 2011; Lu et al. 2017)

$$\cos\theta_v = \cos\theta_{YL} - \frac{\xi v_{CL}(t)}{\sigma_{lv}} - \frac{c_{pin}}{\sigma_{lv}} \operatorname{sgn}(v_{CL})$$

$$c_{pin} = \sigma_{lv}\left|\cos\theta_{eq} - \cos\theta_a\right| \text{ or } c_{pin} = \sigma_{lv}\left|\cos\theta_{eq} - \cos\theta_r\right| \tag{14}$$

Symbols appearing in these equations are explained in the nomenclature. In Equation 14, ‘sgn’ refers to the sign of the contact line velocity. Here $\theta_a$ and $\theta_r$ are the advancing and receding contact angles, respectively. For a PDMS coated electrode, $\xi$ = 0.4 Pa-s, the coefficient of friction and $C_{pin}$ = 0.021 N/m, the coefficient related to contact line pinning (Oh et al. 2010; Annapragada et al. 2011; Lu et al. 2017).

The appearance of the Young-Lippmann contact angle as a boundary condition for the three-phase contact line is justified since the time required to reach this state is much smaller than the timescale of continuous motion (Dwivedi and Muralidhar, 2022).

## 4.5 Initial and boundary conditions

Initial position and boundary conditions of a numerically simulated continuous motion of a 6 µl deionized water are shown in Figure 2. In the geometry of an open-EWOD configuration, a sessile water droplet is placed on a single active electrode, surrounded by air. A horizontal ground wire electrode is placed over the top of the droplet. The effect of the ground wire diameter on the changes in the shape of the air-liquid interface is taken to be negligible as long the wire diameter is small in comparison to the droplet. The horizontal wire electrode geometry is defined by specifying *x* and *y* coordinate values at the top of the droplet. The wire electrode is built as a boundary within the simulator. The simulation is initiated by applying a potential difference between the electrode surface and the horizontal wire, with the wire serving as a ground electrode. The initial equilibrium contact angle before applying the voltage is 104° and is matched with the experimental value. The physical and electrical properties of water and air are available in the software. Simulation involves solving Navier-Stokes equations and tracking the volume fractions of the fluid domains. The three-phase contact line where the two fluid phases meet the wall is provided with the dynamic contact angle model.

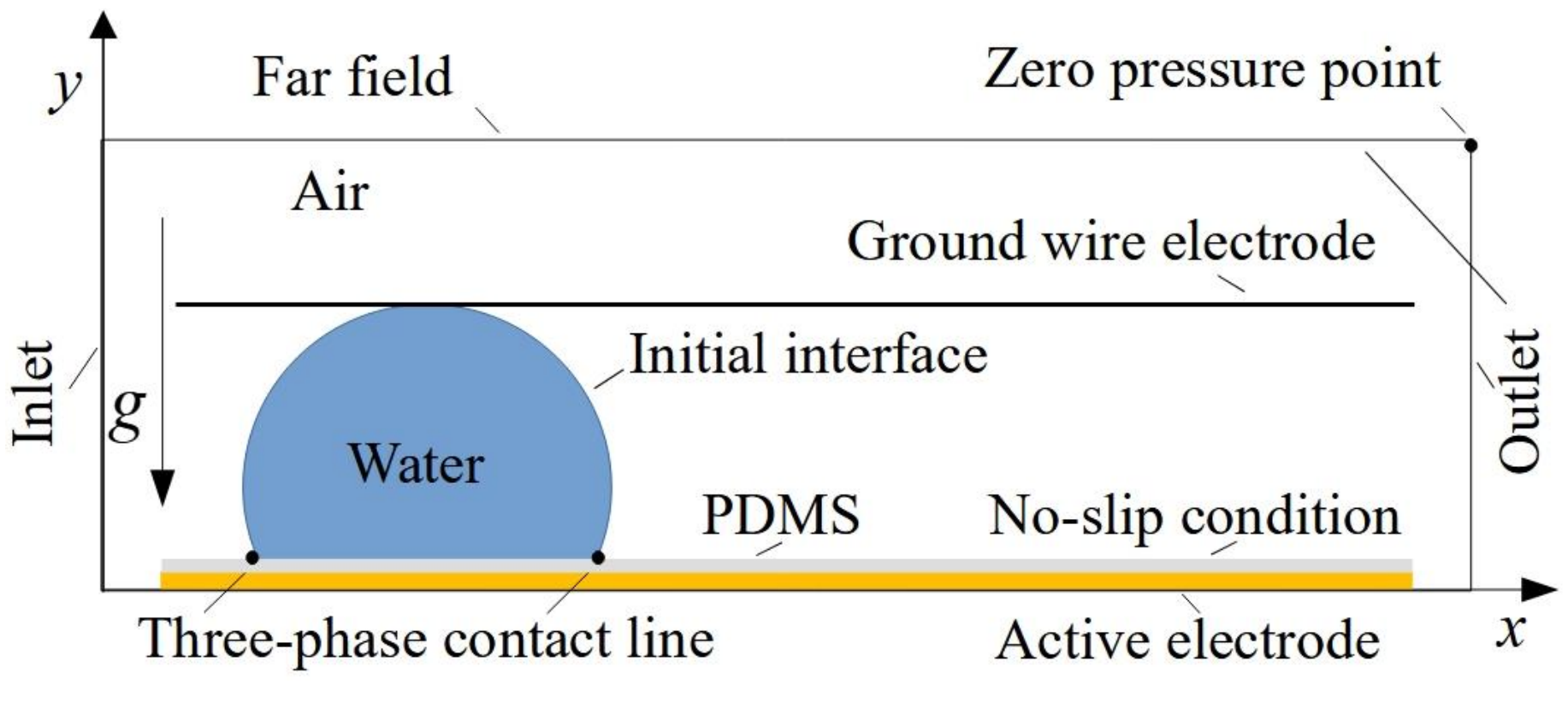

**Figure 2** Boundary conditions of a liquid droplet sitting on a horizontal PDMS-coated copper surface with an equilibrium contact angle of 104° for coupled hydrodynamics-electrostatics simulation

The electrostatic module solves for the Maxwell's stress tensor for the given input voltage. A positive DC voltage is supplied at the horizontal electrode surface and the ground wire connection provided to the wire electrode is zero voltage. The electric field, voltage field, and electrostatic force are outputs that feed the computational fluid dynamics (CFD) simulation. The CFD module solves the Navier-Stokes equations jointly with the phase-field as per the specified interface boundary conditions. The effect of electric field distribution evaluated from the ES module on drop shape and motion is derived in a coupled manner. The dielectric layer in Figure 2 is specified a thickness of 35 µm with the permittivity of PDMS (Wei et al., 2021; Khanna et al. 2020).

In the simulator, air is identified as fluid 1 and water as fluid 2. Symbols $F_x$ and $F_y$ are volumetric force components specified over the entire geometry in the horizontal and vertical directions, respectively; the initial interface defines the boundary between water and air. The constant pressure point is the upper right corner of the air domain. Gravity is included with a built-in gravitational acceleration constant; the reference pressure level is set as one atmospheric pressure, the temperature being 298.15 K (25$^o$C). The initial velocity distribution is zero (quiescent) in all directions. It is necessary to identify parameters of the system that control the speed of the droplet. These are the applied voltage, interfacial tension, liquid viscosity, equilibrium contact angle, contact angle hysteresis, and ground wire position relative to the droplet.

## 4.6 Mesh independence

Various levels of mesh refinement have been considered to ensure adequate resolution of the interfacial thickness and the overall modeling performance. Numerical simulation gradually converges to a grid-independent solution by progressively diminishing the element sizes of the mesh, explicitly, increasing the number of cells. A free triangular mesh is used both on boundaries as well as in internal domains. For drop spreading model, the solution converges at 22 thousand elements with a maximum element size of 78 μm. Hence, the present simulation is performed by using an extremely fine grid that contains a total of 78 thousand elements with a 40 μm maximum element size to enhance the resolution of the interface during spreading. Moreover, for drop motion, the solution converges at 27 thousand elements with an 8 μm maximum element size. Therefore, the present simulation of continuous motion of the drop is performed by using an extra-fine grid that contains a total of 48 thousand elements with a 0.6 μm maximum element size to enhance the resolution of the interface during drop motion. The corner refinement mesh automatically identifies all sharp-enough angles internal to the no-slip boundaries and establishes an even finer mesh along the sharp edge. An increase in cell density is enforced near the walls or sharp edges while maintaining grid quality. In addition, simulations of the present work have been validated against published simulation and experimental data (Section 4.7).

### 4.6.1 Mass balance

In a liquid-air system, the treatment of the interface as diffuse leads to mass loss in the liquid phase while the overall mass of the two-phase system remains conserved. The instantaneous effective mass involved in drop motion and spreading is taken to be the mass displaced from the initial position of the droplet. For the choice of the extra fine mesh and time step, the reduction of the droplet mass is calculated to be 0.47% for the drop spreading model and 0.95% for continuous motion in a coupled electro-hydrodynamic simulation. These errors are small and comparable to those reported in the literature. They confirm the choice of mesh quality and fineness as well as selection of the mobility tuning parameter for simulations of the interface shape, velocity and electric fields. A fine mesh along with a close control on mass balance errors are expected to minimize fluctuations in the interface shape arising from a single-phase formulation of two-phase flow.

## 4.7 Validation study for spreading of a single droplet

The present electrohydrodynamic simulation model of drop spreading dynamics is validated against data reported in the literature. The geometry chosen is axisymmetric. Unsteady data on a DI water drop spreading over a partially wetting dielectric surface in the presence of an electric field generated by a vertical ground wire is chosen as the reference configuration. The present numerical simulation is validated against experiments and ANSYS-FLUENT simulations of Annapragada et al. (2011). Both simulations solve a single set of Navier-Stokes equations and track volume fractions of the fluids assigned in the domains. The friction coefficient and other parameters of the dynamic contact angle are adapted from this reference. Figure 3 shows spreading and contraction of the base radius with a small amplitude of oscillation till nearly a constant base radius is finally attained. The present simulation shows a good match with the literature.

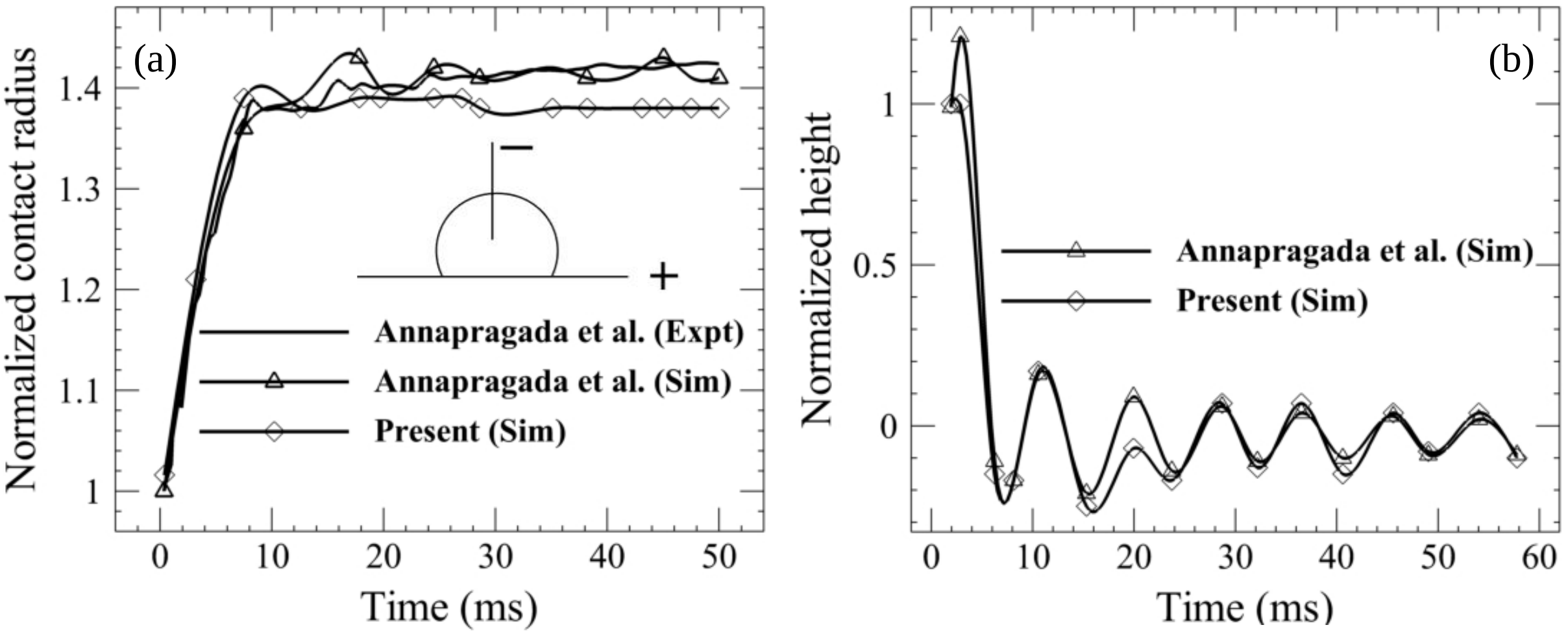

**Figure 3** (a) Comparison of the normalized contact radius between the present numerical simulation against experiments and simulation of Annapragada et al. (2011); (b) time evolution of the normalized height. The present study employs the MKT form of the dynamic contact angle model with contact line friction. The droplet liquid is water of 10 μl volume and the applied voltage is 60 $V_{DC}$.

# 5 Origin of continuous motion of a water droplet

To understand continuous motion of a droplet in an open-EWOD model the analogy with forces acting on a dielectric slab inserted between two parallel-plate capacitors can be proposed (Margulies 1984; Utreras Díaz 1988). The underlying principle can be imagined as the force field on a dielectric slab inserted between two parallel-plate capacitors, namely, the active electrode and the ground wire.

Mugele and Heikenfeld (2019) considered a physical situation similar to the one studied, with the inserted material a perfect conductor and connected the electrodes to a battery. The plate was separated from the electrodes by a thin gap compared to the distance between the two parallel plates. The metallic plate was electrically floating. The electrostatic force acting on the plate was calculated from the gradient of the free electrostatic energy as a function of the plate position. Therefore, the system could be replaced by an electrical equivalent circuit that consisted of a battery and three capacitors representing the parallel plate capacitor without the inserted conducting plate and the two identical gaps between the metal plate and the electrodes. An electrostatic force was exerted on the plate by the electric field, depending mainly on the Maxwell's stresses over the surface. Similarly, liquid droplet placed between the active electrode and the ground wire will experience an initial force when it is placed at one end of the electrode. Here, water drop behaves as a dielectric medium between the active electrode and the ground wire electrode, playing the role of the conducting plate. Once the drop initially deforms, an unbalanced electric field distributed over the air-water interface keeps it in motion. The capacitor analogy is no longer applicable after the drop has left the end of the electrode and is in motion.

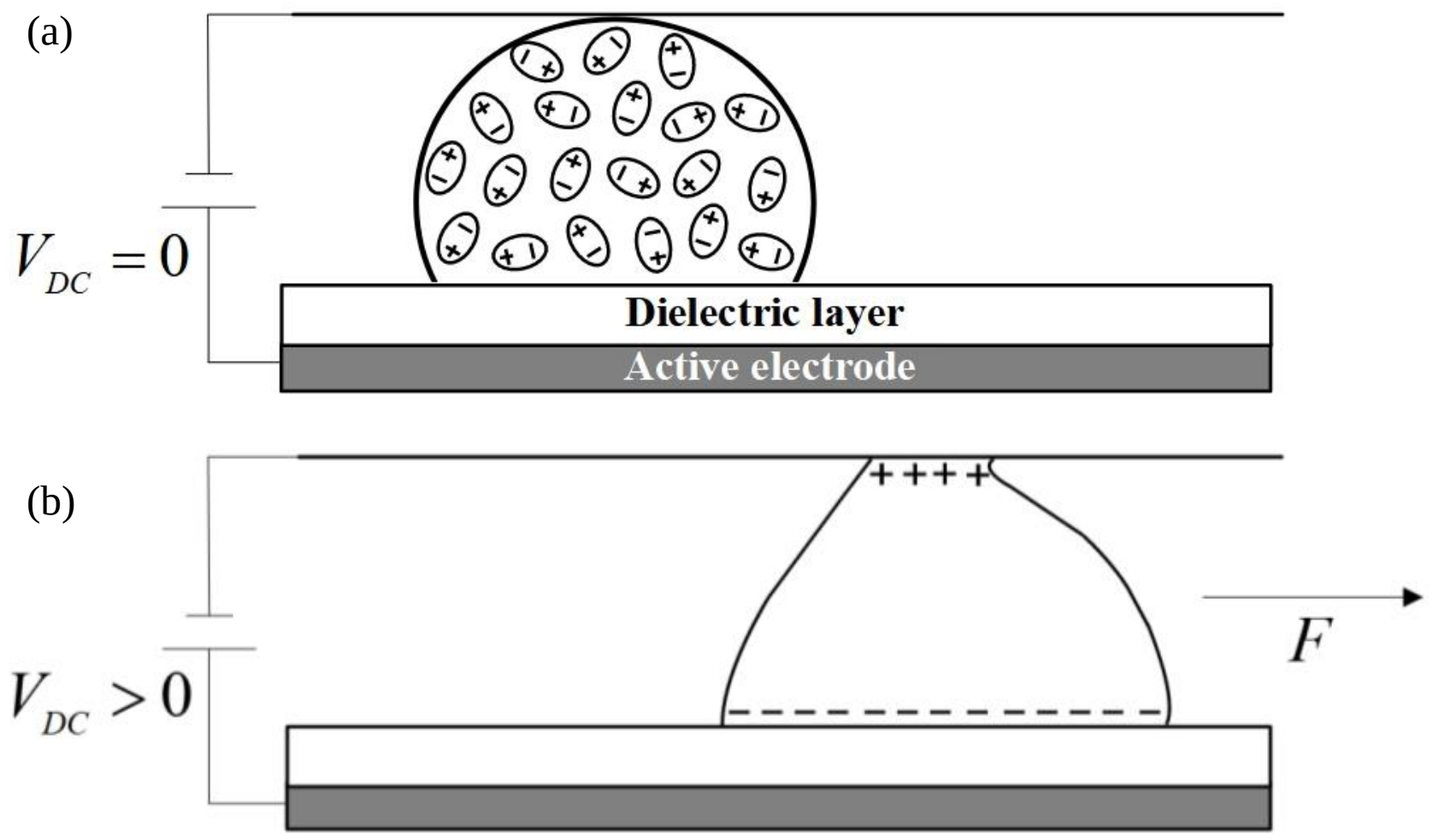

**Figure 4** Schematic representation of surface charge and electric field distribution on a water droplet when voltage is supplied to an open-EWOD device; the water dipoles are indicted by ellipses. (a) Dipoles are randomly distributed at zero voltage when electric fields are not present; (b) Surface charge distribution on a continuously moving droplet. Since the droplet serves as a conductor, the electric field is uniform at the centre of the droplet during motion.

In the open EWOD configuration (Figure 4), the motion of a water droplet is achieved by an unbalanced electrostatic force. Such a situation may arise from a droplet placed initially at the edge of the active electrode. A fringe effect is created at the wedge-like region due to the charge accumulation near the three-phase contact line of the droplet followed by local spreading. On one side of the ground wire, the liquid drop forms an interface with air which is a site for charge accumulation. The interaction of the electric field lines with the water molecules produces the change of shape of the free surface of the droplet. In turn, the electric field is altered as well. Consequently, the drop deforms, creating differences in the three-phase contact angle distribution. These factors will then lead to continuous movement of the drop.

Once motion commences, the bulk of the liquid mass shifts in the direction of motion. The contact angle of the droplet increases on the forward side while decreasing on the rear side. These are respectively the advancing and receding angles of the droplet, and their difference generates an electrical force. Similar changes occur at the ground wire but are harder to characterize owing to the small size of the bridge region. The resultant force unbalance on the droplet arises from a combination of spreading limited by the ground wire position, fringe effect, and finally, the drop deformation due to its forward motion. When the resultant is balanced by the viscous force at the wall, the drop will move at a uniform constant speed parallel to the electrode.

The motion of the drop is possible owing to current drawn from the electrodes. Using an order of magnitude analysis for the drop undergoing steady motion, we may balance the frictional loss at the footprint with the external electric power drawn as follows:

$$\pi d \gamma (\cos\theta_r - \cos\theta_a) \times U = V \times I \qquad (15)$$

Using parameters appearing in the present study, once can estimate the current drawn to be in the range of 0.1-1 μA (10-110 μW) considered in the present study to be negligible. This estimate is the basis for carrying out electrostatic analysis for the force field arising from the charge distribution.

# 6 Results and Discussion

An extensive set of experiments have been carried out to study continuous motion of a water droplet using an open electrowetting on a dielectric system with a DI water droplet sitting on a PDMS-coated active electrode. These include consideration of three different active voltages - 170 $V_{DC,}$ 220 $V_{DC,}$ and 270 $V_{DC}$ for 2 μl, 6 μl, and 10 μl volumes when the ground wire is placed at the top of the droplet. The numerical simulation of a continuous motion of a water droplet is performed. While simulation parameters match experiments, the contact line friction force at the three-phase contact line is set to zero. The effect of non-zero friction is separately considered. The changes in the distribution of the electric field lines arising from the electrode-ground wire geometry as well as drop deformation and motion are the outcome of simulation. Specifically, the ground wire is included in the electric field calculation but is taken to offer negligible resistance to drop motion. Charge accumulation at the three-phase contact line lowers the contact angle leading to spreading. This factor is accounted for by changing the initial contact angle from the equilibrium value to one arising from the Young-Lippmann equation. Next, electrostatic forces are obtained by solving the electric field equations along with suitable boundary conditions in the EWOD geometry. The simulation couples hydrodynamics and electrostatics to jointly determine the instantaneous electric field and droplet motion. The outcome of simulations is compared with experiments when a DC potential difference is created between the droplet and the lower electrode.

## 6.1 Experimental data

The dynamic contact angle of a continuously moving droplet is determined from the extracted images of drop motion under the influence of electric field lines. Specifically, for a drop moving to the right, the angle formed on the right side is the advancing contact angle (ACA), and the one on the left is the receding contact angle (RCA). Both angles are referred to jointly as the dynamic contact angle and are shown in Figure 5. The contact angles arising from the motion of 2, 6, and 10 μl droplets at various operating voltages have been determined using image processing operations from the video sequence of drop motion. For a 2 μl droplet, the change in advancing and receding contact angles is not significant for 170 and 220 $V_{DC}$ voltages. However, a notable difference from the initial equilibrium condition is realized for an operating voltage of 270 $V_{DC}$. Part of the reduction is related to spreading arising from electrowetting of the liquid drop over the dielectric and is proportional to the square of the voltage. The rest of the increase arises from an altered pressure distribution and flow within the drop.

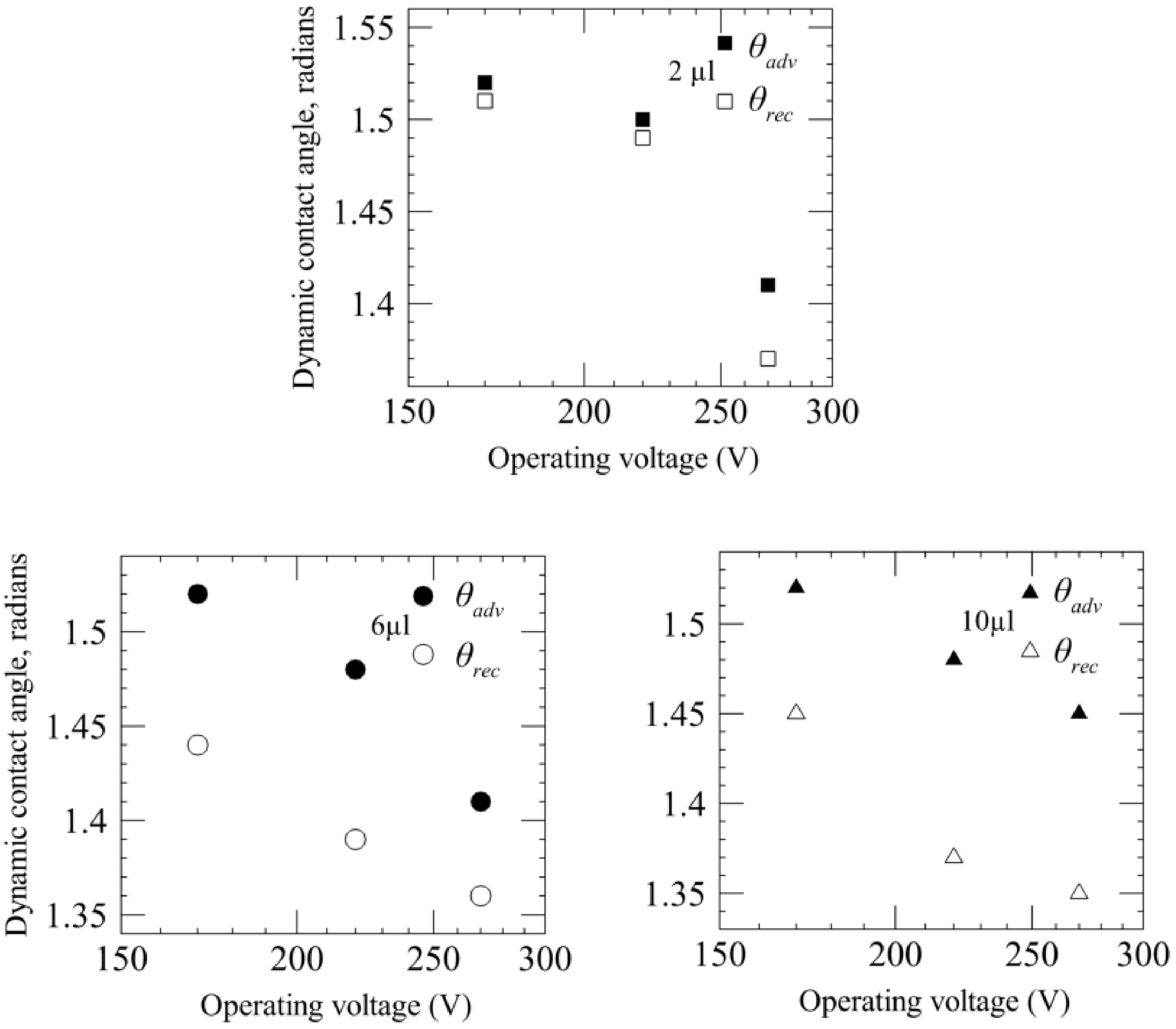

**Figure 5** Experimentally determined advancing and receding contact angles attained during electrically driven continuous motion of a water droplet as a function of the operating voltage and droplet volume

A substantial influence of an increase in operating voltage is seen during the motion of 6 µl and 10 µl droplets. The change is more significant when the voltage increases from 220 to 270 $V_{DC}$ compared to an increase from 170 to 220 $V_{DC}$. Inertial effects increase with liquid volume and contribute to changes in the contact angle over and above what is attained by electrowetting alone. Specifically, inertia alters surface curvature and hence pressure distribution and increases the advancing angle relative to equilibrium while further decreasing the receding contact angle.

The terminal velocity attained by the droplet in terms of the capillary number as a function of the excitation voltage is shown in Figure 6. It is to be expected that the role of inertia forces will increase continuously with the droplet volume. Under steady velocity conditions, frictional resistance at the wall will balance contact line force arising from variation in the contact angle along the footprint, while the inertia force itself goes to zero. Both factors will increase with velocity. Figure 6 shows that the velocity attained by the 10 µl droplet is generally higher than the other two volumes since a higher surface area will generate a higher unbalanced electrical force. Speed increases with applied voltage, but the dependence is not linear. The change from 220 to 270 $V_{DC}$ is marked when compared to a change from 170 to 220 $V_{DC}$. It should be remarked here that the voltages that contribute to motion are those available after accounting for the polarization of the PDMS layer serving as a dielectric.

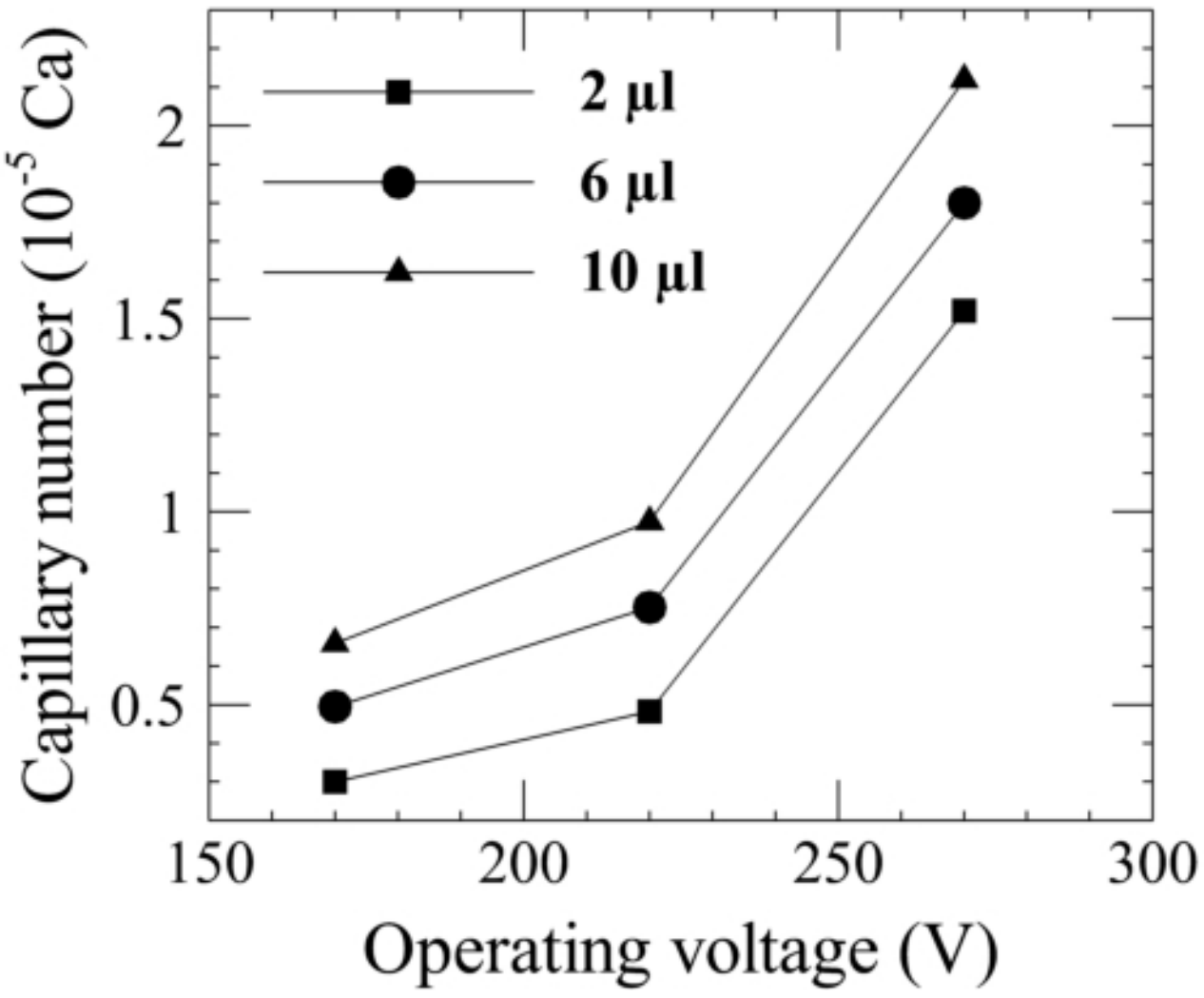

**Figure 6** Experimentally determined increase in capillary number as a function of the operating voltage for various volumes of the droplet.

The deformation of the interface, reduction of wetting angle on the solid surface, and pulling of the liquid toward the central part of the electrode are achieved by electrostatic actuation. Larger droplets accelerate faster than smaller droplets since the interaction area of the droplet with the electric field increases as the droplet volume increases (Figure 6). The resulting motion of the water droplet is governed jointly by the electric field and the hydrodynamic field. As the magnitude of the electric field increases, the water droplet cannot maintain a stable state, resulting in droplet deformation and motion. The contact line remains pinned below the threshold voltage, namely, the smallest voltage needed to initiate droplet motion. However, the drop starts to deform, losing symmetry in the process. Once motion is initiated, the electrical field is partially balanced by viscous and pressure forces, the difference leading to droplet acceleration.

The left-right symmetry is initially broken by placing the drop on the left side of the electrode, allowing it move to right. If the drop were to be placed on the right, symmetry would be broken in reverse and the drop would move to the left. The droplet was placed at the left extreme to make it move towards right. Hence, this initial condition does break symmetry and initiates motion.

For a constant potential difference between the electrode and the ground wire, a water droplet in contact with a hydrophobic electrode surface tends to spread, deform and end in continuous motion. The force arising from the electric field has two components in the horizontal and vertical directions, the latter balanced by the normal reaction at the electrode. The horizontal component mobilizes the drop, generates unequal advancing and receding angles, and is finally balanced by viscous friction at the wall and a resultant interfacial force distribution that opposes motion. Spreading is related to charge accumulation at the three-phase contact line (to be called Young-Lippmann spreading), in turn, constrained by the position of the ground wire that enforces contact with the liquid at all times. In the final stages, the change in the shape of the drop is small, and it may be taken to have achieved a terminal velocity. At this stage, inertia force reduces to zero.

For the geometry studied, the electrode is of finite dimensions, and the drop reaches its other end. Here, the drop decelerates and finally comes to rest. In the following discussion, the phases of motion starting from rest, acceleration, and attainment of constant velocity alone are considered. The forces that lead to drop movement are local differences in the contact angle around the three-phase contact line (or contact points, in a two-dimensional setting) and distribution of the force arising from the electric field around a slightly asymmetric drop. These are resisted by viscous drag at the electrode that increases linearly with drop velocity and a component of interfacial tension at the air-liquid interface. Under conditions of steady motion, the sum of all external forces is zero. Hence, one may conclude that continuous motion of a DI water droplet is achieved over a single active electrode in three stages: static → spreading; spreading → acceleration; acceleration → steady motion. The related free body diagram of the drop at each stage of motion is presented in Figure 7.

The static position of the droplet sitting over the hydrophobic deactivated electrode surface is shown in Figure 7(a) when there are no external horizontal and vertical forces present. A vertical downward arrow indicates the force of gravity due to the mass of the drop. The upward arrows represent the force exerted across the liquid-solid interface. The tangential components of the interfacial tensions act along the solid-liquid-air three-phase contact line. Pressure shown is the gage value relative to the atmosphere. As soon as voltage is applied over the electrode, the drop starts spreading under the influence of an electric field. Both contact angle as well as the shape of the droplet change (Figure 7(b)). At the same time, the presence of electric field lines over the air-liquid interface modifies its shape. The change in contact angle of the droplet due to the applied voltage is shown in Figure 7(c). Correspondingly, $F_{YL}$ is the pseudo force acting at the three-phase contact line for the given voltage. Once maximum spreading of the droplet is achieved, the drop deforms and accelerates, as shown in Figure 7(d). The vectorial sum of forces - electrostatic ( $F_E$ ), viscous ( $F_v$ ), contact line ( $F_{CL}$ ) and surface tension ( $F_{ST}$ ) acting in the horizontal direction is non-zero. When the sum is positive, the motion of the droplet is initiated.

Terminal velocity by the droplet is achieved when the sum of all forces in the horizontal direction is zero while the shape of the drop remains unchanged during its motion, Figure 7(e).

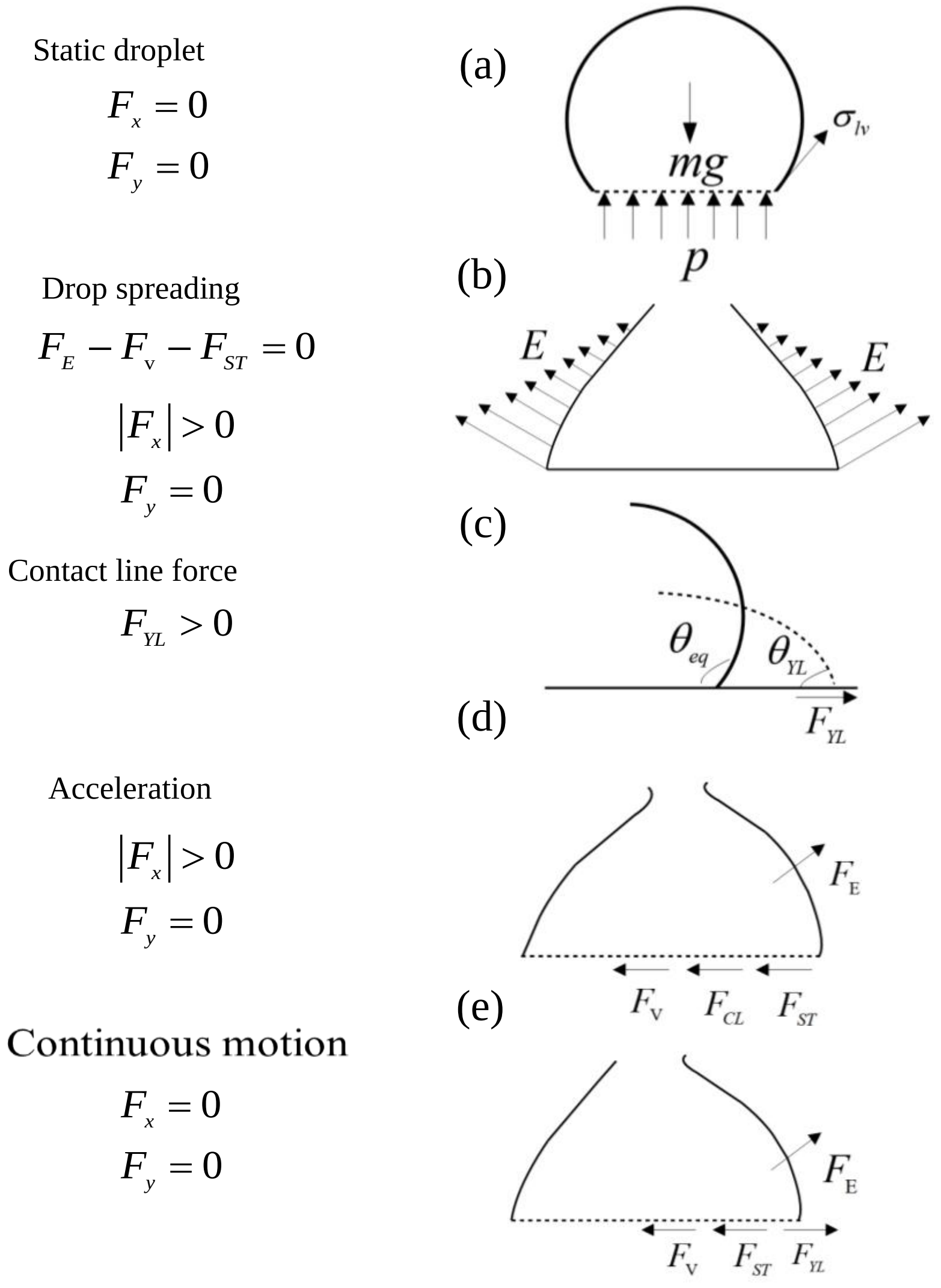

**Figure 7** Free body diagram of the liquid drop during various stages of its motion: (a) static shape, (b) instantaneous Young-Lippmann spreading, (c) unbalanced force at the three-phase contact line; (d) bulk acceleration; and (e) steady continuous motion

## 6.2 Comparison of experiments with simulation

A qualitative comparison between experiments and simulations in terms of interfacial shapes of a drop moving in an electric field is discussed. The experimental and simulation image sequence of droplet motion over a bare PDMS surface and silicone oil-coated PDMS surface at 270 $V_{DC}$ is shown in Figs. 8 and 9, respectively, extracted from the experimental and simulation image sequence. The shapes change rapidly during the acceleration stage. The changes are larger for a higher potential difference. In both experiments and simulations, a stable drop shape is attained during continuous motion with constant velocity.

### 6.2.1 Friction and pinning effect of PDMS dielectric layer

A difference observed in drop speeds between experiments and simulation over bare PDMS and silicone oil coated PDMS may be attributed primarily to surface characteristics, particularly pinning and friction. A PDMS coating over the electrode behaves as a sticky hydrophobic surface that serves to resist droplet motion. This issue was addressed in the present study in two different ways. In an experiment, silicone oil (μ = 5 mPa-s) was coated over the PDMS dielectric layer, and the drop actuation experiments were repeated. Silicone oil rendered the PDMS layer as a lubricated hydrophobic surface, lowering the coefficient of friction and possibly eliminating pinning. In another approach, simulation was improved by including a dynamic contact angle (Equation 14) that explicitly accounted for the friction of the dry PDMS layer (Oh et al. 2010; Annapragada et al. 2011; Lu et al. 2017). With both of these corrections, the match between experiments and simulation (Figs. 8 and 9) was seen to have substantially improved. Figures 8 and 9 show that reduced friction increases the terminal velocity attained by the drop.

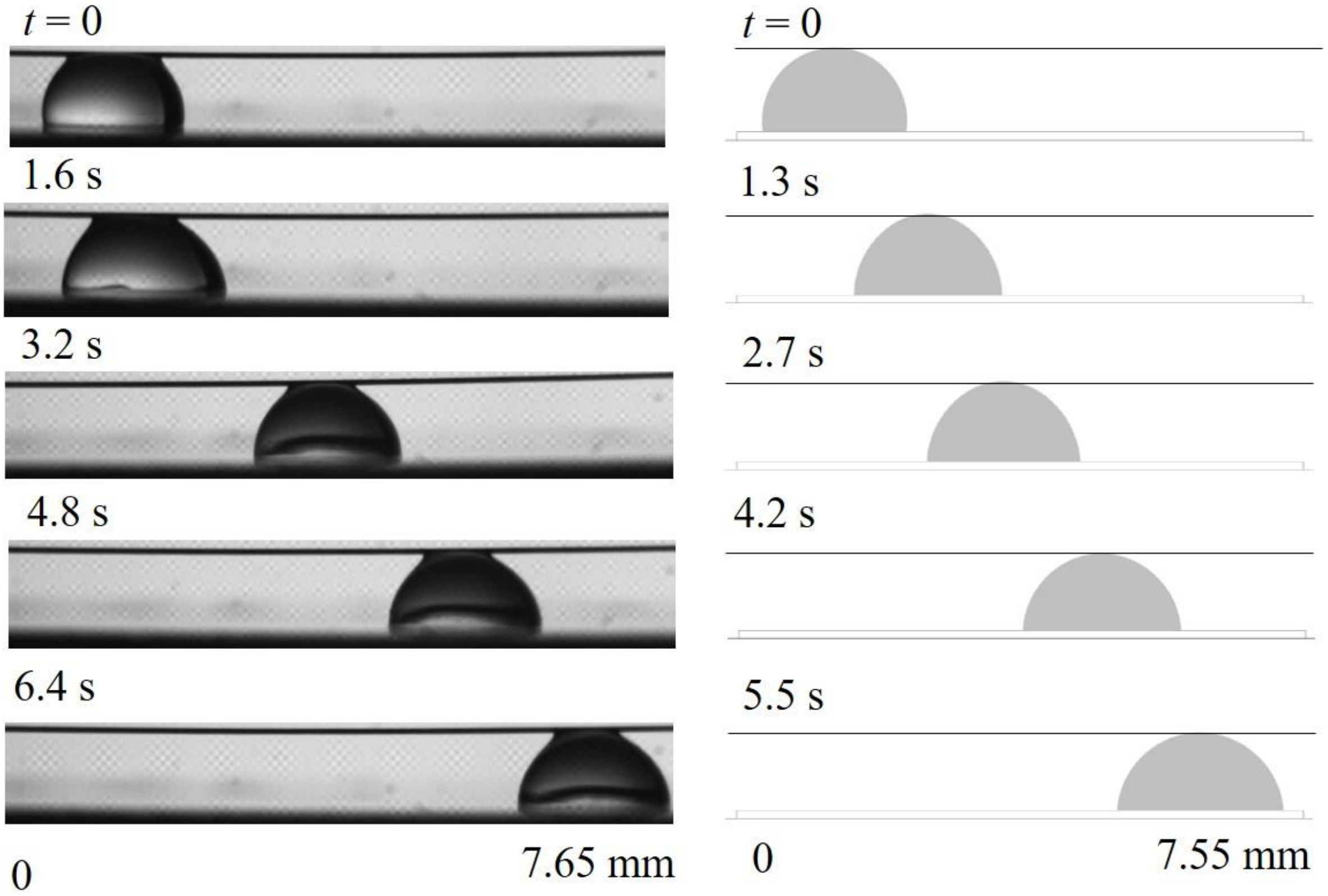

**Figure 8** A continuously moving 6 μl water droplet at 270 $V_{DC}$: (a) experimental image sequence of droplet motion over a bare PDMS surface; (b) numerically simulated image sequence of droplet motion arising from the DCA model with friction and pinning. The droplet boundary is defined at the interface ($V_{f,2}$ = 0.5) between the water droplet (gray) and surrounding air (white).

A video of drop motion over a silicone oil coated PDMS material is presented as supplementary material (video SI).

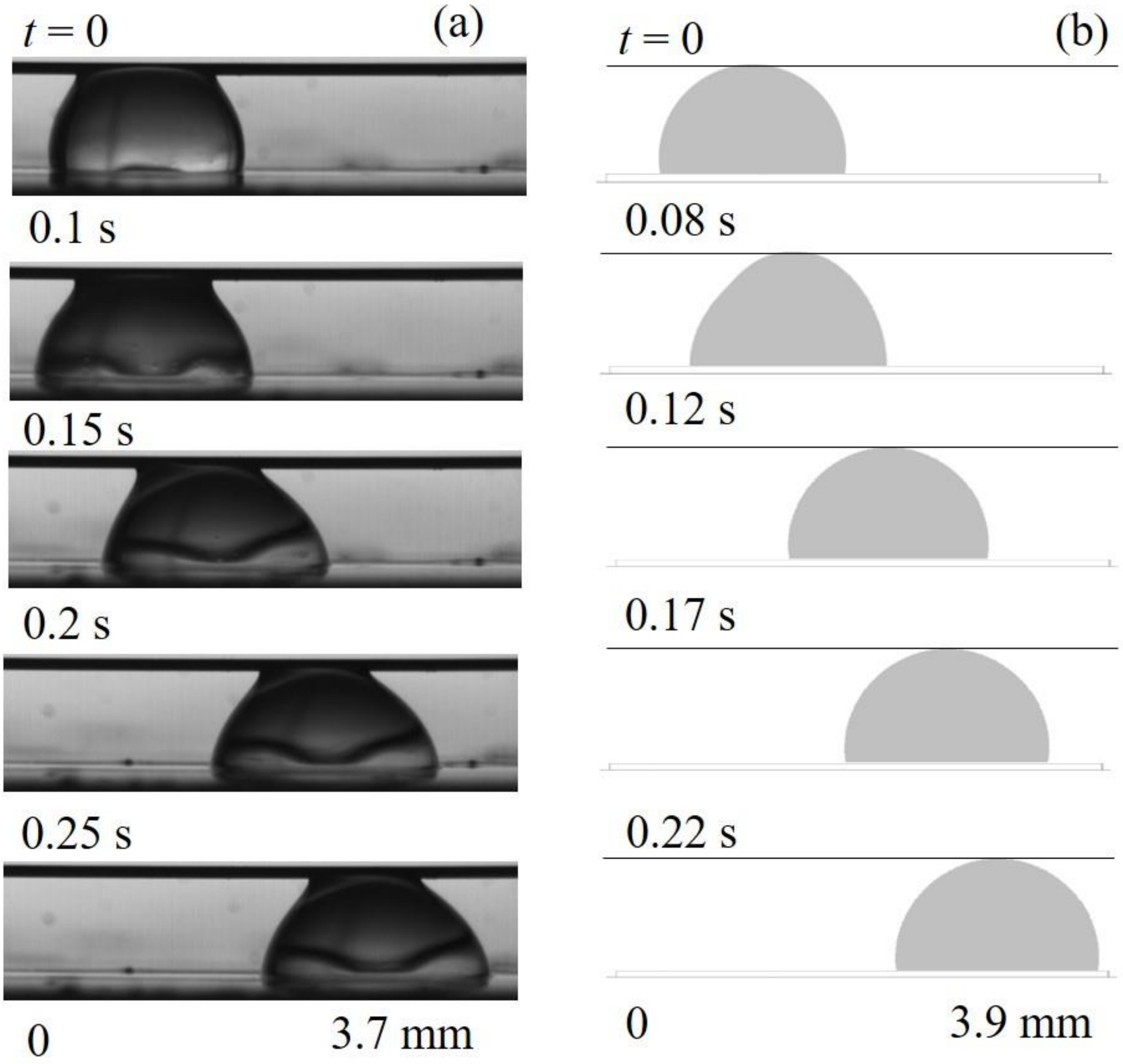

**Figure 9** A continuously moving 6 µl water droplet at 270 $V_{DC}$: (a) experimental image sequence of droplet motion over the silicone oil-coated PDMS surface; (b) numerically simulated image sequences of droplet motion arising from the CCA model without friction. The droplet boundary is defined at the interface ($v_{f,2}$ = 0.5) between the water droplet (gray) and surrounding air (white).

### 6.3 Velocity and pressure field

Numerically simulated velocity vectors in and around the droplet are shown in Figs. 10(a) and 11(a) at selected instants of time. Figure 10(a) includes contact line friction in the dynamic contact angle model. Figure 11(a) is drawn from a simulation with a constant dynamic contact angle, which is equivalent to zero friction at the three-phase contact line. The figures are quite similar except for a change in the time scale, with wall friction slowing down the drop movement.

### 6.3.1 Relative velocity vectors

The average $x$-component velocity within the droplet at a given time instant is calculated by extracting $x$-component velocity at 500 points within the droplet and then averaging them. The average $x$-component velocity within the droplet is then determined using the following expression:

$$u_{avg} = \frac{1}{n}\sum_{i=1}^{n} u_i \tag{16}$$

Here, $i = 1...n$ is a number of points tracked inside the droplet to find the average. The relative $x$-component velocity is found by subtracting the average $x$-component velocity value from the local value at a given time instant. Hence

$$u_{rel} = u_i - u_{avg} \tag{17}$$

Simulations showed that the average $y$-component velocity was close to zero. The relative velocity vectors plotted within the water droplet are shown in Figs. 10(b) and 11(b). A circulation pattern is clearly seen in the acceleration stage of the drop and weakens with time as the drop approaches a steady state.

Figures 10 and 11 reveal three stages of drop motion, specifically early spreading, acceleration, and finally a steady speed. Initially, the droplet starts spreading when a potential difference is created between the wire and active electrode surface. Subsequently, the droplet accelerates, leading finally to continuous motion with nearly a constant speed. Figures 10(b) and 11(b) show vectors of the relative velocity drawn with respect to a frame of reference moving at the spatially-averaged velocity of the liquid phase. Here, it is seen that droplet deformation is the highest during the acceleration regime of drop movement while the relative velocity magnitudes are also substantial. The acceleration phase supports a recirculation pattern within the drop that moves forward with it. As the drop speed approaches a constant terminal value, the relative velocity magnitudes diminish and the circulation pattern fades away. In terms of the absolute values, velocities in Figs. 10(a) and 11(a) straighten up in the terminal velocity regime, indicating steady translation of the drop between the two electrodes. Drop deformation is seen to be the highest during the acceleration stage, particularly during motion over a lubricated PDMS surface.

### 6.3.2 Total pressure

Internal pressure distribution at selected instants of time within the drop during spreading, acceleration, and steady motion are shown in Figs. 10(c) and 11(c). These figures jointly show interface movement of the drop and a grey-scale distribution of pressure. In these figures, the atmospheric pressure is the datum and a dark shade is positive. Thus, pressures within the drop are positive. Interface pressures are affected by local curvature on one hand and the unbalanced electrical force, the two factors aiding each other at some locations and opposing each other at others. In the absence of an electric field, regions of low curvature will correlate to low internal pressure when compared to high curvature where the pressure is higher. This difference will show up as a jump in pressure across the interface. In the electrowetting context, spreading over the active electrode lowers the jump discontinuity. Subsequently, the drop is set in motion, which redistributes pressure on each side of the interface. Figs. 10(c) and 11(c) show that the internal pressure in the drop continues to be high for all stages of its motion, a result consistent with the fact that the drop integrity is maintained during its motion. However, at the interface itself, the pressure discontinuity is weakened owing to the electric field and is further modulated by the drop deformation and inertia in the spreading, acceleration, and steady stages of its movement. The greatest departure in the pressure-curvature relationship occurs near the apex of the drop, where the small diameter of the ground wire enhances the electric field and increases further the local interface curvature. Elsewhere, pressure is correlated with the local instantaneous curvature of the drop. Large droplet deformation and negative curvature is to be seen at a time instant of 180 ms (Figure 11(c)) in numerical simulation over a silicone oil-coated surface during the acceleration stage, for which higher velocities are attained. Such locations will serve as sites of negative pressure and hence, possible instability that may result in the splitting of the drop. However, during the spreading and steady stage, curvature is everywhere positive, stabilizing the drop shape. In contrast, experiments (Figs. 8 and 9) did not show regions of negative curvature during drop motion. Here, it is to be borne in mind that simulations are in two dimensions while experiments concern full droplets that prevent the formation of any zone of negative curvature. Additionally, Figs. 10 and 11 do not account for the physical presence of a ground wire in hydrodynamic simulations, and the possibility of instability is high. The influence of a ground wire on drop shapes in discussed later (Figure 14) and is shown to improve drop stability characteristics.

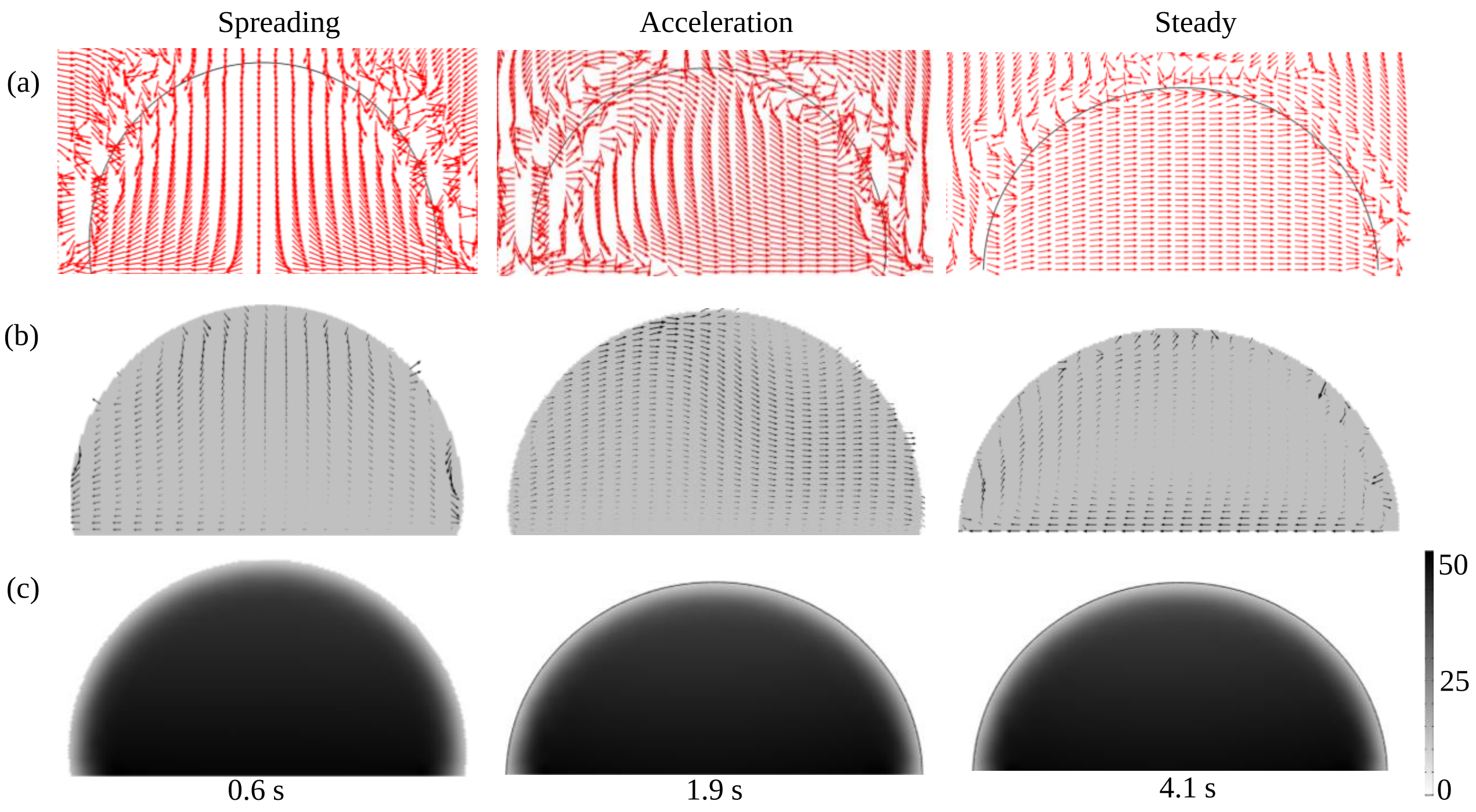

**Figure 10** Three regimes of continuous motion for a 6 µl droplet at 270 $V_{DC}$ over a bare PDMS surface: (a) velocity vectors in and around the droplet at various time instants; (b) relative velocity vectors where the maximum arrow length in *x*-direction corresponds to the positive velocity of 0.08 m/s; (c) Total pressure distribution under the influence of an electric field. The dynamic contact angle model in simulations includes friction and pinning

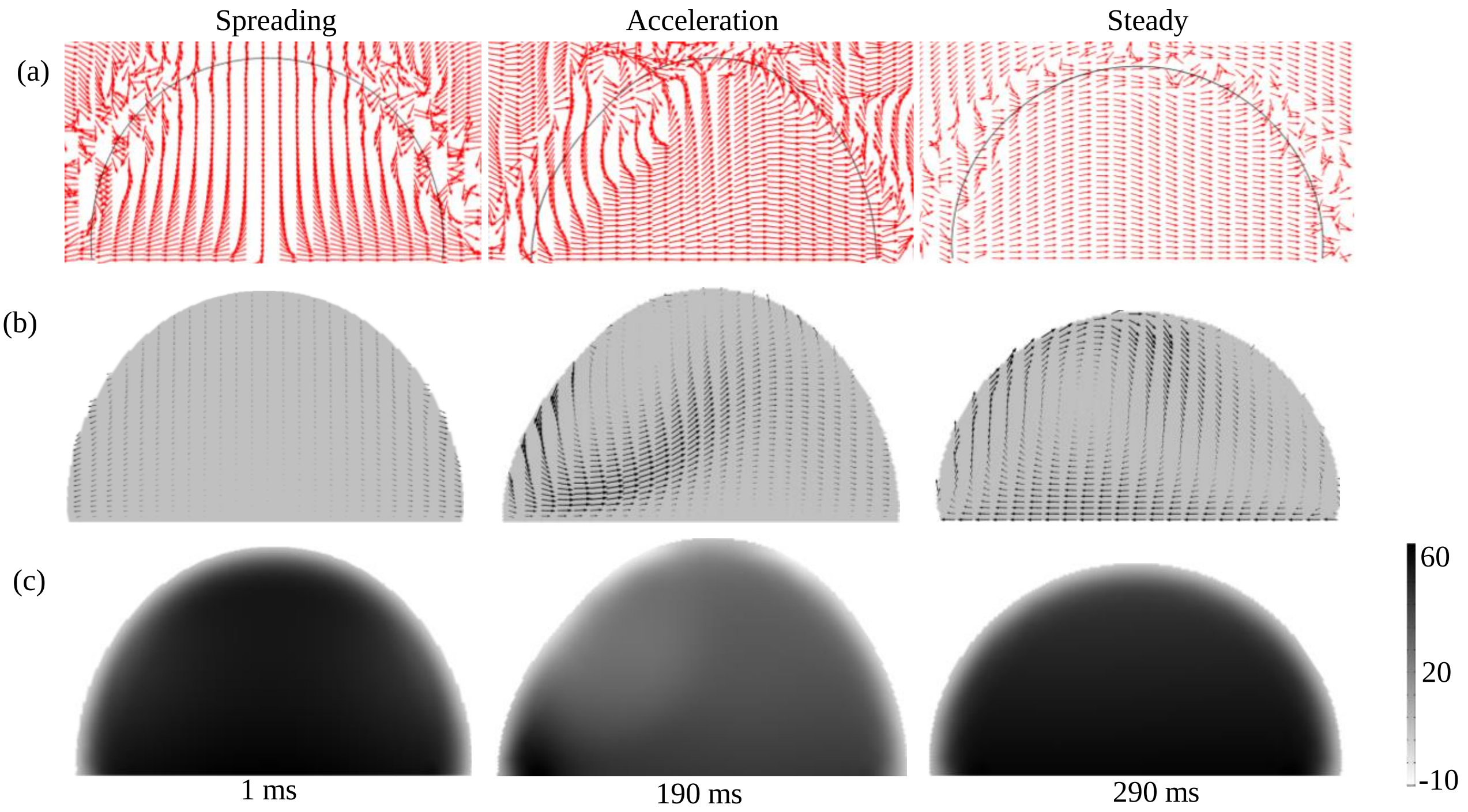

**Figure 11** Three regimes of continuous motion for a 6 µl droplet at 270 $V_{DC}$ over a silicone oil coated surface: (a) velocity vectors in and around the droplet at various time instants; (b) relative velocity vectors where the maximum arrow length in *x*-direction corresponds to the positive velocity of 0.17 m/s; (c) Total pressure distribution under the influence of an electric field. The dynamic contact angle model in simulations neglects friction and pinning.

## 6.4 Electrostatic force field vectors at the interface

The electrostatic force field is vectorially represented in Figure 12 with and without contact line friction. The resultant unbalanced force in the positive *x*-direction shows the actuation component that drives the drop forward. The vertical component adds on the contribution of pressure as well as gravity, and is oppositely directed. Figure 12 shows that the *x*-component of the electric force is the highest during the acceleration phase and is related primarily to asymmetry in drop deformation. The two effects are coupled and can only be realized in a coupled simulation. The resultant force in the direction of movement diminishes as the velocity attains a constant value.

The local electric field intensity at the interface between the droplet and the surrounding fluid arises from the permittivity contrast of air and water at the interface. Apart from the external voltage, the magnitude of the permittivity contrast determines charge accumulation and hence, the polarization force acting at the interface.

The force field distributed over the interface is shown in Figure 12(b) for a silicone oil-coated PDMS surface without friction. The drop starts spreading from the initial condition when the voltage is applied ($t$ = 0) to attain a maximum contact radius at $t$ = 120 ms. Consequently, the resultant electrostatic force field vector was obtained in the forward *x*-direction. The length of the arrow is proportional to the strength of the electrostatic force. Afterward, the drop deforms and the acceleration stage commences at $t$ = 170 ms where the electrostatic force field strength is large. Finally (~ 300 ms), steady motion of the droplet is attained when the electrostatic force field is reduced, making the unbalanced external force zero. Thus, the three stages of the drop motion, namely, spreading, acceleration, and steady motion correspond closely to the electrostatic force field vector distribution over the air-water interface.

The differences between the electrostatic force field vectors at the interface with and without contact line friction at the three-phase contact line are not noticeable. However, a slightly larger deformation of the droplet is observed in Figure 12(a) since friction and pinning at the contact line restrict the deformation of the droplet. The resultant electrostatic force pushes the droplet forward but also impacts the normal reaction at the lower surface. This influence is seen as a disturbance in the wall pressure distribution over the droplet footprint.

### 6.4.1 Equipotential lines

For a droplet sitting over a horizontal electrode with a ground wire over it, droplet motion is initiated first in the form of spreading, followed by acceleration till the drop is set in uniform motion. The equipotential lines for these three regimes are discussed in the present section. The drop behaves essentially as a conductor and the accompanying voltage drop is small.

Soon after the voltage is turned on, the drop starts spreading under the influence of the electric field. This trend is seen in both experiments and simulations. Spreading is related to charge accumulation at the three-phase contact line and is a local effect. Hence, the symmetry of the droplet is maintained during spreading. Spreading continues as long as the drop maintains contact with the ground wire and ceases at later times. Simultaneously, the drop develops a certain amount of asymmetry, both because of the presence of the ground wire as well as its position over the active electrode. The forces at the liquid-air interface respond to this asymmetry, and a net force acts on the droplet in the forward direction. The drop starts movement over the electrode and continues to deform till a limiting shape is attained. The liquid within the droplet begins to shift forward as a result of inertial forces, and a clear difference between the advancing and receding angles is visible. Thus, the three regimes of drop movement - spreading, acceleration, and uniform velocity are demarcated in terms of differences in the advancing and receding angles.

Figure 13 depicts the change of shape of the droplet with the passage of time. The transient shape of the droplet and change in the equipotential lines during the continuous motion of the droplet is shown for the 6 µl droplets at 270 $V_{DC}$. A reduction in the spacing of the equipotential lines signifies a higher field intensity, higher gradients, and hence, larger force components. It confirms that the electrostatic forces are distributed over the liquid-air interface.

The electrostatic forces increase with an increase in the electric field intensity, resulting in higher deformation in such regions. With reference to Figure 13(a) and (b), note that $t$ = 1.4 s in Figure 13(a) and $t$ = 5 ms reveals that the droplet has deformed and shifted to the right in response to the non-uniformity in the electric field. Thus, the drop starts moving to the right. The coupling of electric forces, drop deformation, and motion is thus established.

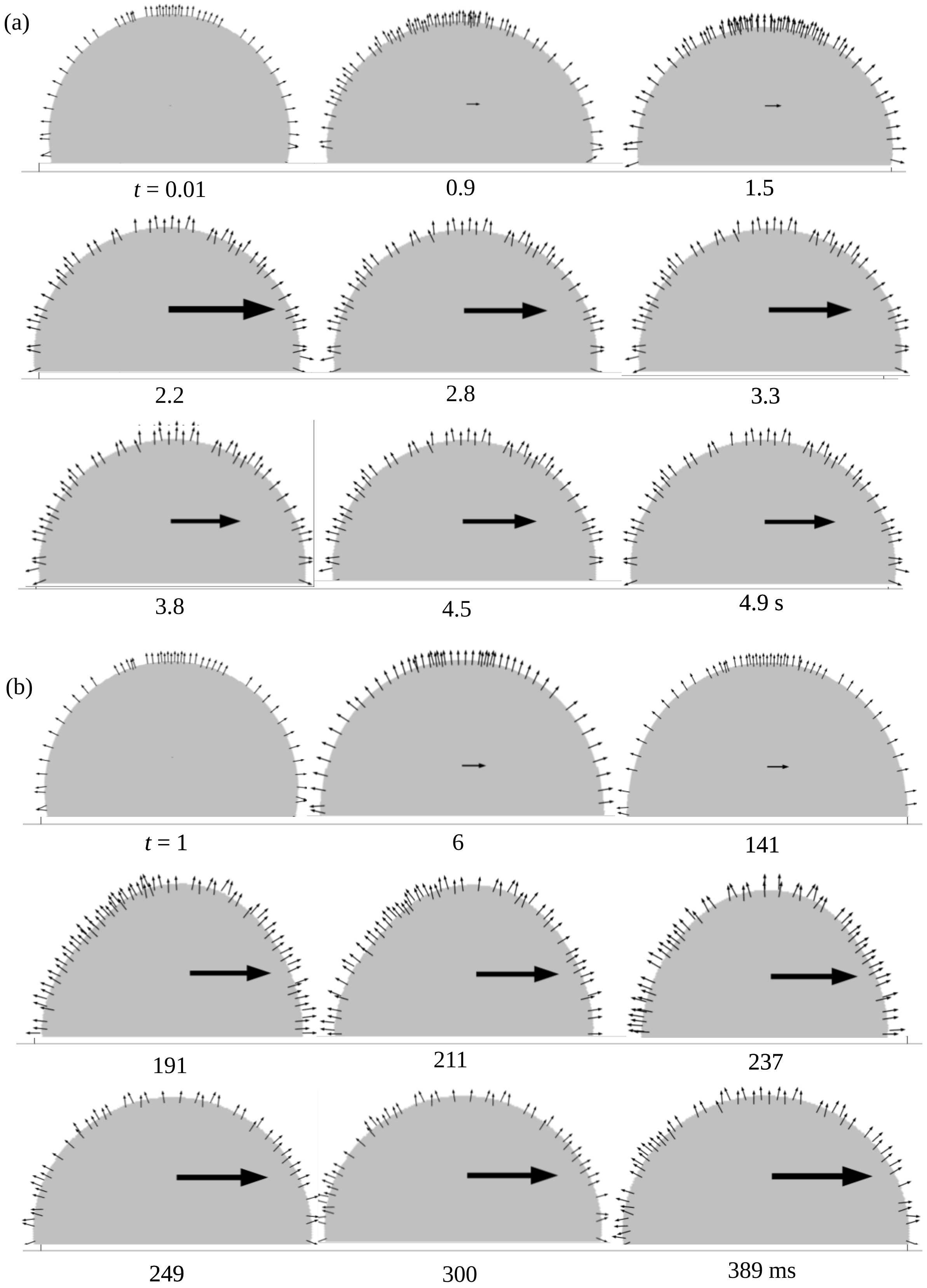

**Figure 12** The electrostatic force field vectors distributed over the water-air interface and the resultant electrostatic force in the *x*-direction in the center of a 6 µl droplet for an applied voltage. Simulation of drop motion arising from the (a) DCA model with friction and pinning and (b) CCA model without friction

.

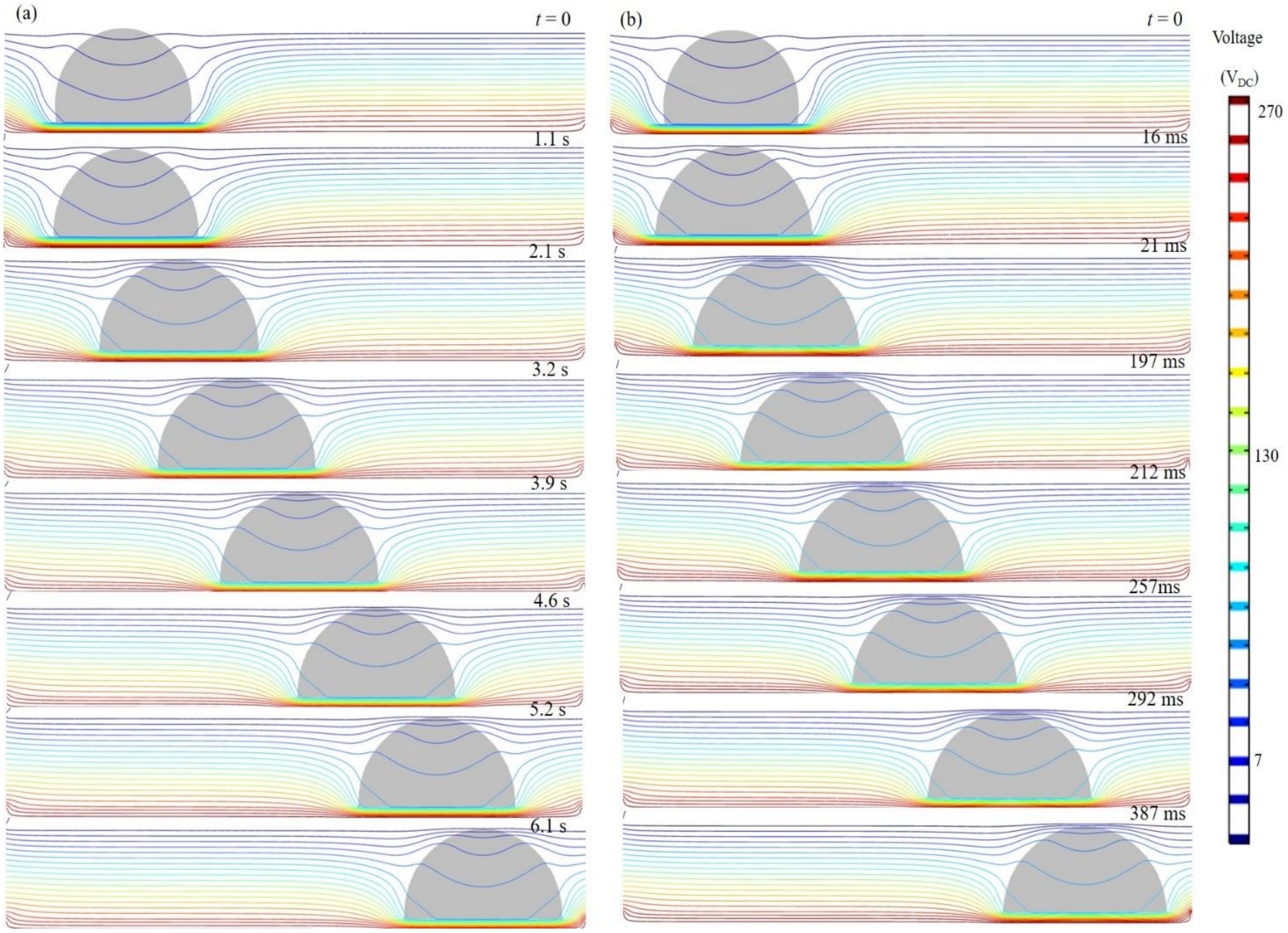

**Figure 13** Displacement of the simulated equipotential lines between the active electrode (270 $V_{DC}$) and the ground wire during motion of a 6 µl water droplet. The equipotential lines are mainly parallel to the active and ground electrodes and distortion is restricted to the air-water interface. (a) DCA model with friction and pinning; (b) CCA model without friction.

## 6.5 Wettability of the ground wire

In order to have a complete understanding of the effect of the wettability of the ground wire over the deformation of the droplet under the influence of the electric field, selected numerical simulations were carried out. The surface considered is silicone oil-coated PDMS, droplet volume is 6 µl, and the applied voltage is 270 $V_{DC}$. Here, the wettability effect of the ground wire is factored in the hydrodynamic module of the simulator. For a hydrophilic wire, the drop spreads over it, diminishes its local curvature, and increases deformation as shown in Figure 14. The present simulation (Figure 14(b)) thus shows better match with the experimental drop shapes of Figure 14(a).

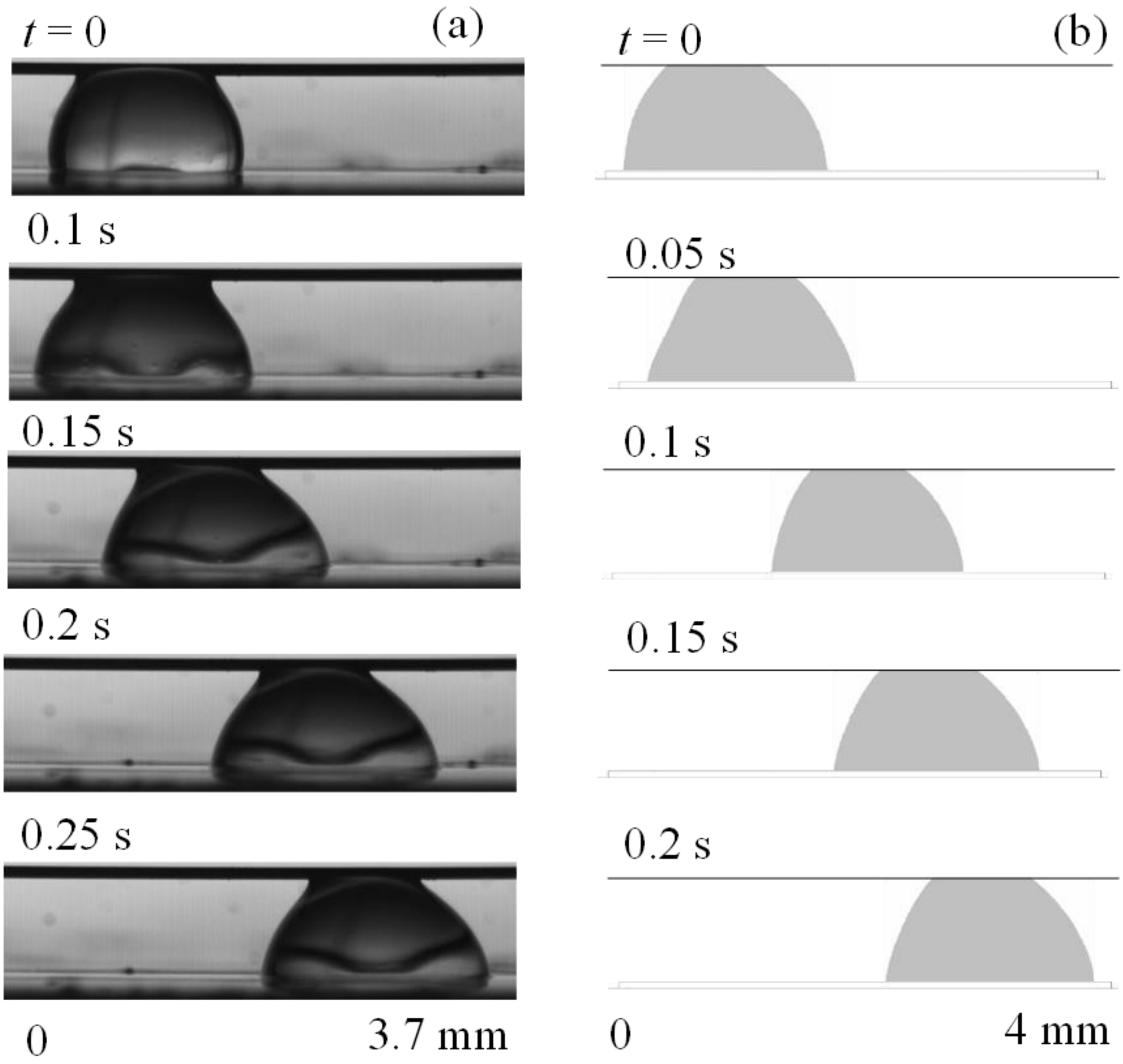

**Figure 14** A continuously moving 6 µl water droplet at 270 $V_{DC}$: (a) image sequence of droplet motion over the silicone oil-coated PDMS surface recorded in experiments; (b) numerically simulated image sequence of droplet motion arising from the CCA model without friction. The physical presence of the ground wire is included in the hydrodynamic simulation coupled with electrostatics

## 7 Conclusions

Continuous motion of a DI water droplet over a PDMS-coated single electrode is achieved using an open-EWOD arrangement with a horizontal ground wire placed just above. Apart from spreading, the role of the electric field is to create asymmetry in the shape of the drop, thus leading to its continuous motion over the electrode. The asymmetry in the shape of the droplet is connected to non-uniformity in the electric field, thus coupling the electrostatic and hydrodynamic forces. The largest unbalanced force distribution occurs at the interface where there is a jump in electrical permittivity.

Droplet motion has been analyzed numerically in a coupled electro-hydrodynamic simulation with a dynamic contact angle model that includes electro-wetting, friction and hysteresis. These simulations match experiments of electrically actuated continuous motion of a drop over bare PDMS and silicone oil-coated PDMS surfaces.

Continuous motion of a water droplet is seen to be achieved experimentally over a single active electrode in three stages: spreading, acceleration, steady motion. These stages are reproduced in simulations. The three stages of drop motion can be correlated with velocity, pressure, and electric field distributions. During acceleration, velocity vectors show the formation of a circulation pattern that moves with the droplet. On a lubricated surface, velocity attained is higher, the drop deformation is higher, and the pressure field is considerably disturbed. Excess pressure in the drop shows a direct relationship with the interface curvature and changes sign when the interface goes from being convex to concave. The skewness in the electric field lines provides an explanation for the electrostatic force acting on the drop.

The distance between the horizontal ground wire from that electrode placed parallel to it is an important parameter. For a wire placed initially at the top of the drop, this constraint ensures that spreading as well as the droplet asymmetry is limited, and the droplet speed does not increase drastically with external voltage. Since the overall drop deformation is limited, the advancing and receding angles attain nearly equal values. With an increase in volume, the surface area also increases, and the electric field deforms the drop to a greater extent. Thus, higher velocity is achieved for a larger volume of the droplet and with an increase in the applied voltage.

The importance of the wettability of the ground wire over the deformation of the droplet is revealed in simulations. For a hydrophilic wire, the drop spreads over it, diminishes its local curvature, and increases deformation.

**Supplementary information**

The submitted files include a video (supplementary information SI).

**Declarations**

**Ethical Approval**
Not applicable

**Competing interests**
None declared.

**Authors' contributions**
The work is drawn from the doctoral dissertation of the first author for which the second author served as the thesis guide.

**Funding**
Authors acknowledge fellowship received by the first author from the Ministry of Education, Government of India.

**Availability of data and materials**
Raw data can be made available from the authors on reasonable request.

**Conflict of interest** There are no conflicts to declare.

**Nomenclature**

| | |
|---|---|
| $d$ | Thickness of the dielectric layer, µm |
| $E$ | Electric field, V/m |
| $E_x$ | Horizontal components of the electric field, V/m |
| $E_y$ | Vertical components of the electric field, V/m |
| $F_E$ | Electrostatic force, N/m$^3$ |
| $F_x$ | Horizontal components of the volumetric electrostatic force, N/m$^3$ |
| $F_y$ | Vertical components of the volumetric electrostatic force, N/m$^3$ |
| $F$ | Body force, N/m$^3$ |
| $F_G$ | Gravitational force, N/m$^3$ |
| $F_{ST}$ | Interfacial force of surface tension, N/m$^3$ |
| $F_{ext}$ | External body force field, N |
| $F_{\mathrm{v}}$ | Viscous force, N/m$^3$ |
| $N$ | Number of pixels |
| $n$ | Number of points |
| $p_{avg}$ | Average wall pressure, N/m$^2$ |
| $\overleftrightarrow{T} = \mathrm{T}_{ij}$ | Maxwell stress tensor, N/m$^2$ |
| $u$ | x-component velocity of the droplet, m/s |
| $v_{CL}(t)$ | Contact line velocity, m/s |
| $V$ | Operating voltage, V |
| $V_{DC}$ | Direct current voltage, V |
| $W$ | Weight factor for pixels |
| $w$ | Area function |
| $x_c$ | Non-dimensional *x*-coordinate of centroid |
| $y_c$ | Non-dimensional *y*-coordinate of centroid |

**Greek symbols**

| | |
|---|---|
| $\gamma$ | Mobility parameter, m$^3$·s/kg |
| $\varepsilon_o$ | Permittivity of free space, F/m |
| $\varepsilon_d$ | Permittivity of the insulating dielectric layer |
| $\varepsilon_r$ | Relative permittivity of the material |
| $\varepsilon_a$ | Relative permittivity of air |
| $\varepsilon_w$ | Relative permittivity of water |
| $\varepsilon_{pf}$ | Parameter controlling interface thickness, m |
| $\xi$ | Friction factor, Pa-s |
| $\theta_{adv}$ | Advancing contact angle, degrees |
| $\theta_{rec}$ | Receding contact angle, degrees |
| $\theta_d$ | Dynamic contact angle, degrees |
| $\theta_{eq}$ | Equilibrium contact angle, degrees |
| $\theta_{e,dyn}$ | Dynamic contact angle (Annapragada et al. 2011) |
| $\theta_{YL}$ | Contact angle obtained from Young–Lippmann equation, degrees |
| $\theta_v$ | Contact angle under the influence of applied voltage, degrees |
| $\lambda$ | Mixing energy density, N |
| $\mu$ | Viscosity, N-s/m$^2$ |
| $\mu_w$ | Viscosity of water, N-s/m$^2$ |
| $\mu_a$ | Viscosity of air, N-s/m$^2$ |
| $v_{f,1}$ | Volume fraction distribution of surrounding fluid |
| $v_{f,2}$ | Volume fraction distribution of droplet |
| $\rho$ | Density, kg/m$^3$ |
| $\rho_l$ | Density of water, kg/m$^3$ |
| $\rho_a$ | Density of air, kg/m$^3$ |
| $\sigma$ | Surface tension coefficient at the water-air interface, N/m |
| $\sigma_{lv}$ | Surface tension coefficient between liquid-vapour, N/m |
| $\phi$ | Phase-field variable |